\documentclass[aps,superscriptaddress,amsmath,amssymb,twocolumn,showpacs,floatfix,english,noshowpacs]{revtex4}

\usepackage{url}
\usepackage{bm,bbm}
\usepackage{graphicx}
\usepackage{color}
\usepackage[colorlinks=true, urlcolor=blue, linkcolor=blue, citecolor=blue, pdftex]{hyperref}
\usepackage[english]{babel}

\usepackage{soul}
\usepackage{blindtext}
\usepackage{lipsum}
\usepackage{nicefrac}

\newcommand{\dagga}{{\phantom{\dagger}}}
\begin{document}

\title{Charge-density waves in kagome-lattice extended Hubbard models \\ at the van Hove filling}

\author{Francesco Ferrari}
\affiliation{Institute for Theoretical Physics, Goethe University Frankfurt, Max-von-Laue-Stra{\ss}e 1, D-60438 Frankfurt a.M., Germany}
\author{Federico Becca}
\affiliation{Dipartimento di Fisica, Universit\`a di Trieste, Strada Costiera 11, I-34151 Trieste, Italy}
\author{Roser Valent\'\i}
\affiliation{Institute for Theoretical Physics, Goethe University Frankfurt, Max-von-Laue-Stra{\ss}e 1, D-60438 Frankfurt a.M., Germany}

\date{\today}

\begin{abstract}
The Hubbard model on the kagome lattice is presently often considered as a minimal model to describe the rich low-temperature 
behavior of AV$_{3}$Sb$_{5}$ compounds (with A=K, Rb, Cs), including charge-density waves (CDWs), superconductivity, and possibly broken 
time-reversal symmetry. Here, we investigate, via variational Jastrow-Slater wave functions, the properties of its ground state when both onsite 
$U$ and nearest-neighbor $V$ Coulomb repulsions are considered at the van Hove filling. Our calculations reveal the presence of different interaction-driven CDWs 
and, contrary to previous renormalization-group studies, the absence of ferromagnetism and charge- or spin-bond order. No signatures of chiral 
phases are detected. Remarkably, the CDWs triggered by the nearest-neighbor repulsion possess charge disproportionations that are not compatible 
with the ones observed in AV$_{3}$Sb$_{5}$. As an alternative mechanism to stabilize charge-bond order, we consider the electron-phonon interaction, 
modeled by coupling the hopping amplitudes to quantum phonons, as in the Su-Schrieffer-Heeger model. Our results show the instability towards a 
tri-hexagonal distortion with $2\times 2$ periodicity, in a closer agreement with experimental findings.
\end{abstract}

\maketitle

{\it Introduction.}
The interplay between electronic correlation and lattice geometry is the source of several fundamental phenomena in condensed-matter systems.
The Hubbard model, with nearest-neighbor hopping amplitude $t$ and onsite repulsion $U$~\cite{hubbard1963}, represents the simplest way to 
describe interacting electrons in a crystal. The competition between the kinetic processes and the Coulomb interaction is enhanced on frustrated 
lattices, giving rise to a rich physical behavior. In two dimensions, a particularly interesting case is represented by the kagome lattice, for 
which a few studies have focused on the Mott transition at half filling~\cite{bulut2005,ohashi2006,kuratani2007,guertler2014,sun2021,kaufmann2021}. 
Extended Hubbard models on the kagome lattice involving the nearest-neighbor interactions $V$ have been explored as well, such as the case of 
spinless fermions at $1/3$ filling, where the Fermi energy lies at the Dirac points of the non-interacting band 
structure~\cite{nishimoto2010,obrien2010,wen2010,ruegg2011,ferhat2014}. Within the spinful case, some attention has been given to the model at 
$5/6$ filling, where the Fermi energy intersects a Hove singularity~\cite{yu2012,kiesel2012,kiesel2013,wang2013}. This scenario is particularly 
interesting because the Bloch states connected by the nesting vectors display different sublattice characters, thus obstructing the onset of 
electronic instabilities generated by the Hubbard-$U$ interaction, which acts on the same sublattice~\cite{kiesel2012}. Therefore, the 
nearest-neighbor $V$ term can play an important role. Indeed, renormalization-group analyses have shown the appearance of several unconventional 
electronic phases, including ferromagnetism and charge- or spin-bond orders, although considerably different phase diagrams have been obtained 
by two independent calculations~\cite{kiesel2013,wang2013}. 

A renovated interest in the properties of the kagome lattice system at the van Hove filling has been sparked by the recent discovery of the family 
of AV$_{3}$Sb$_{5}$ metals (with A=K, Rb, Cs)~\cite{ortiz2019}. Their {\it ab initio} electronic band structure displays different van Hove 
singularities in the proximity of the Fermi energy, originating from the $d$-orbitals of Vanadium atoms, which form almost perfect two-dimensional 
kagome layers. Upon lowering temperature, AV$_{3}$Sb$_{5}$ materials undergo two subsequent transitions~\cite{jiang2021_rev}, first developing 
charge-density wave (CDW) order in an intermediate regime~\cite{jiang2021,zhao2021,uykur2021,ortiz2021,liang2021,shumiya2021,li2022}, and then 
exhibiting superconductivity at lower temperatures~\cite{ortiz2020,ortiz2021b,yin2021,chen2021}. The CDW phase requires a $2\times 2$ supercell 
within the Vanadium layers~\cite{jiang2021,zhao2021}, with star-of-David and/or tri-hexagonal patterns~\cite{christensen2021,hu2022}. Interestingly,
different experimental probes have detected signatures of time-reversal symmetry breaking in the CDW phase, stimulating the understanding of its
origin~\cite{yang2020,jiang2021,shumiya2021,wang2021,wu2021,yu2021,mielke2022}.

The starting point of most theoretical studies is the Hubbard model on the kagome lattice with a single orbital per site, originating from the 
$d_{xy}$-orbitals of the Vanadium atoms~\cite{jiang2021,denner2021,jeong2022}. Within this minimal formulation, different CDW phases have been 
proposed to arise as potential instabilities of the electronic band structure at the van Hove filling, some of them featuring non-trivial orbital 
currents~\cite{park2021,lin2021,feng2021,feng2021b,mertz2022,yang2022}. The microscopic physical mechanism triggering the CDW instability is still 
under debate. While an early mean-field analysis has indicated the nearest-neighbor electronic repulsion as the possible origin of the chiral CDW 
observed in AV$_{3}$Sb$_{5}$~\cite{denner2021}, several works suggest that lattice deformations and electron-phonon coupling may play an important 
role~\cite{wenzel2021,xie2021,ratcliff2021,tan2021,luo2022,wu2022,liu2022,mei2022}. 

Motivated by these studies, we revisit the problem of the extended Hubbard model on the kagome lattice at the van Hove filling, with $U$ and 
$V$ terms. We employ a variational Monte Carlo approach based on Jastrow-Slater wave functions to map out the phase diagram of the model and 
analyse the CDW instabilities induced by the electronic repulsion. Our results show that different CDWs can be stabilized in the phase diagram,
but no ferromagnetism is present. In addition, charge- or spin-bond order is detected only within uncorrelated states (i.e., without the Jastrow 
factor), while the correlated Jastrow-Slater wave functions do not show any evidence for this kind of instabilities. Our outcomes are in striking 
contrast to previous calculations based on functional renormalization group~\cite{kiesel2013,wang2013}. Most importantly, charge modulations 
generated by the Coulomb repulsion $V$ display a substantial disproportionation on neighboring sites, which is not compatible with the $2\times 2$ 
CDW observed experimentally~\cite{jiang2021,zhao2021,uykur2021,ortiz2021,liang2021,li2022}, where the electron density retains an almost perfect
$C_6$ rotational symmetry around the center of hexagons (forming star-of-David or tri-hexagonal patterns). For this reason, we also analyze the 
effect of the electron-phonon coupling on the Hubbard model (without $V$), where phonons affect hopping amplitudes, as in the Su-Schrieffer-Heeger 
model~\cite{su1979}. In this case, lattice distortions appear, with short bonds along disconnected hexagons, favoring a charge reorganization 
that is similar to the one observed in AV$_{3}$Sb$_{5}$~\cite{christensen2021,hu2022}. 

{\it The purely electronic model.}
We consider the (extended) Hubbard model for spinful electrons on the kagome lattice 
\begin{eqnarray}\label{eq:hubham}
\mathcal{H}&=&-t \sum_{\langle i,j \rangle,\sigma} (c^\dagger_{i,\sigma} c^\dagga_{j,\sigma} + H.c.) \nonumber \\
&+& U \sum_{i} n_{i,\uparrow} n_{i,\downarrow} + V \sum_{\langle i,j \rangle} n_i n_j,
\end{eqnarray}
where $t>0$ is the nearest-neighbor hopping term, and ${U\geq 0}$ and ${V\geq 0}$ denote the strength of the onsite and nearest-neighbor repulsive 
interactions, respectively. The fermionic operator $c_{i,\sigma}$ ($c^\dagger_{i,\sigma}$) annihilates (creates) an electron with spin $\sigma$ on 
site $i$. The Coulomb interactions are expressed in terms of the number operators $n_{i,\sigma}=c^\dagger_{i,\sigma}c^\dagga_{i,\sigma}$ and 
$n_i=n_{i,\uparrow}+n_{i,\downarrow}$. The total number of sites in the system is denoted by $N$, while the total number of electrons is 
$N_{\rm e}=N_\uparrow+N_\downarrow$, with $N_\sigma=\sum_i n_{i,\sigma}$. In the following, we focus on the filling $n_F=N_{\rm e}/N=5/6$, for 
which the non-interacting Fermi energy intersects the upper van Hove singularity~\cite{supp_mat}.

We investigate the Hamiltonian of Eq.~\eqref{eq:hubham} by a variational approach, which relies on the use of Jastrow-Slater wave functions to 
approximate the ground state of the Hubbard model, and Monte Carlo sampling to stochastically compute observables. Our variational {\it Ans\"atze} 
take the general form $|\Psi_{\rm e}\rangle=\mathcal{J} |\Phi_0\rangle,$ in which a long-range density-density Jastrow factor
\begin{equation}
\mathcal{J}=\exp\left(\sum_{i,j} v_{i,j} n_i n_j\right)
\label{eq:Jastrow}
\end{equation}
is applied on top of an uncorrelated fermionic state, $|\Phi_0\rangle$, to introduce non-trivial correlations between electrons. The resulting 
wave function goes beyond standard mean-field approaches based on uncorrelated states, and can potentially describe different phases of Hubbard-like
models~\cite{capello2005,capello2006,tocchio2013,tocchio2014,kaneko2016,bijelic2018}, e.g., metallic and (Mott or band) insulating phases.
We take the uncorrelated part of the variational state, $|\Phi_0\rangle$, to be the ground state of an auxiliary tight-binding Hamiltonian 
\begin{equation}\label{eq:mfham}
\mathcal{H}_0 = -\sum_{\langle i,j \rangle,\sigma} T^{\sigma}_{i,j} (c^\dagger_{i,\sigma} c^\dagga_{j,\sigma} + H.c.) - \sum_i \mu_i c^\dagger_{i,\sigma} c^\dagga_{i,\sigma},
\end{equation}
featuring nearest-neighbor hopping terms ($T^{\sigma}_{i,j}$) and onsite potentials ($\mu_i$). 
The parameters of $\mathcal{H}_0$ are optimized 
together with the (translationally invariant) $v_{i,j}$ parameters defining the Jastrow factor, to minimize the variational energy of 
$|\Psi_{\rm e}\rangle$. The optimization is performed numerically by means of the stochastic reconfiguration technique~\cite{sorella2005}.
Further details on the variational calculations are reported in the Supplemental Material~\cite{supp_mat}.

\begin{figure}
\includegraphics[width=\linewidth]{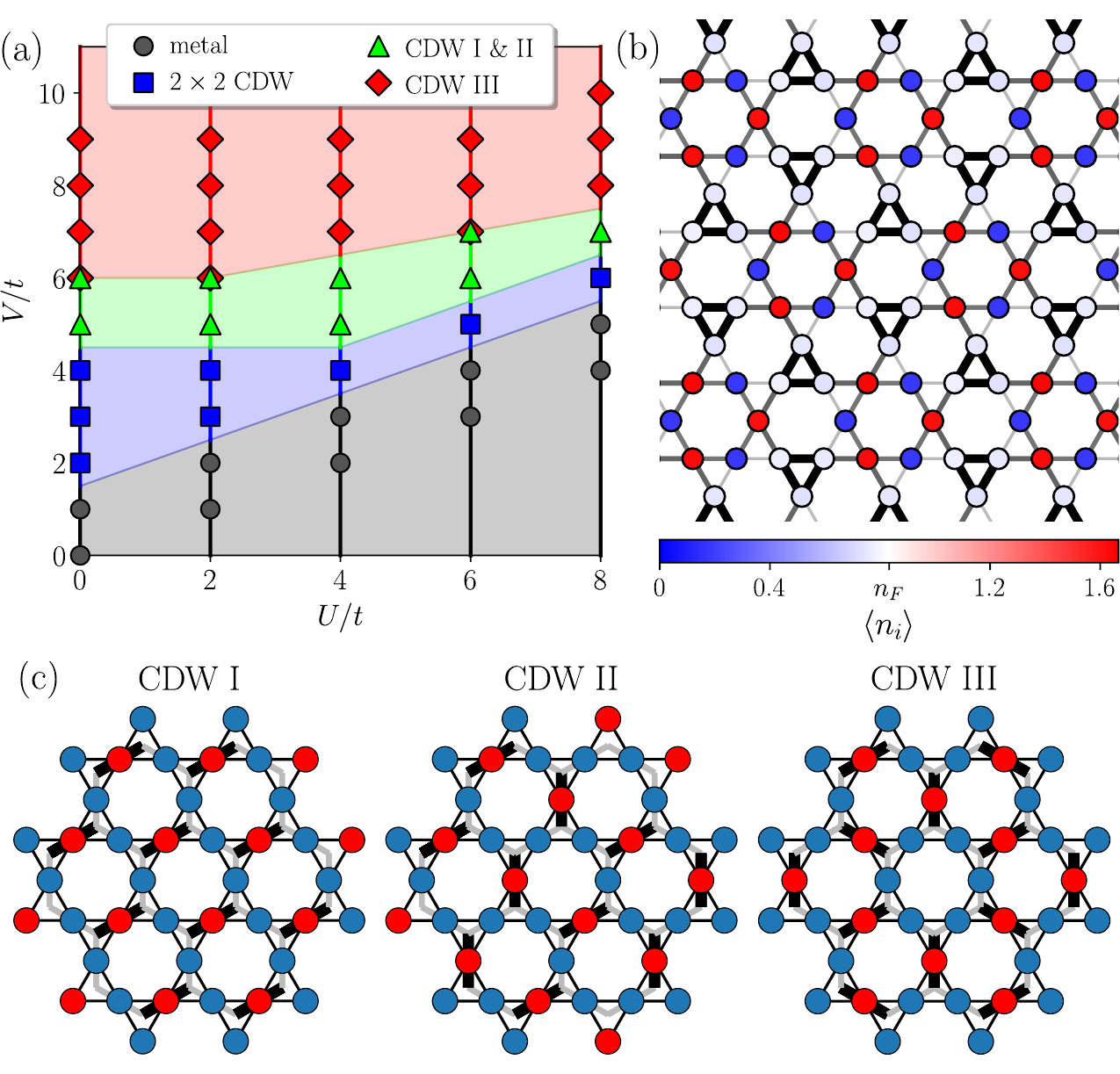}
\caption{\label{fig:phasediag}
Panel (a): phase diagram of the Hubbard model of Eq.~\eqref{eq:hubham} at the van Hove filling $n_F=5/6$. The symbols indicate the values of $U/t$ 
and $V/t$ for which the calculations have been performed (on finite-sized lattices with $L_1=12$, and $L_2=10$ and $18$~\cite{supp_mat}). 
Overlapping symbols are used when the variational energies of two phases are equivalent within errorbars. Panel (b): charge pattern in the 
$2\times 2$ CDW phase. The color of each site $i$ indicates the average number of electrons $\langle n_i \rangle$, while the width and the darkness 
of the lines connecting nearest neighbors $(i,j)$ are proportional to the modulus of the expectation value of the hopping operator along the bond, 
namely $|{\langle c^\dagger_{i,\uparrow} c^{\protect\dagga}_{j,\uparrow} +c^\dagger_{i,\downarrow} c^{\protect\dagga}_{j,\downarrow} \rangle}|$. 
Results for $U/t=0$ and $V/t=3$ are shown. Panel (c): sketch of the electronic charge patterns of the CDW I, II, and III phases, fulfilling the 
{\it triangle rule}. Blue (red) circles denote depletion (accumulation) of electrons. Following Ref.~\cite{obrien2010}, for each CDW we show the 
dimer configuration on the honeycomb lattice formed by connecting the centers of the corner-sharing triangles.}
\end{figure}

{\it Results.}
The phase diagram of the model, as obtained by our variational approach, is shown in Fig.~\ref{fig:phasediag}~(a), featuring a metallic phase 
for (relatively) small values of $V/t$, and different CDW phases, illustrated in Fig.~\ref{fig:phasediag}~(b)-(c), which are stabilized by the 
presence of a sizeable nearest-neighbor interaction.

The variational {\it Ansatz} for the metallic phase is obtained by applying the Jastrow factor $\mathcal{J}$~\eqref{eq:Jastrow} on top of the 
uncorrelated ground state of the uniform auxiliary Hamiltonian of Eq.~\eqref{eq:mfham} with $T_{i,j}^\sigma=1$ at nearest-neighbors, and $\mu_i=0$ 
for all sites. The resulting wave function, which reduces to the exact ground state of the model in the non-interacting limit ($U=V=0$), turns out 
to provide the optimal variational energy in the whole metallic phase  [grey region in Fig.~\ref{fig:phasediag}~(a)]. No sign of ferromagnetism is 
observed in the metallic region, in constrast to previous renormalization group results~\cite{kiesel2013,wang2013}. Indeed, by performing the 
variational calculations for different values of the magnetization $m=(N_\uparrow-N_\downarrow)/N$, the minimal energy is always found to be at 
$m=0$ (while the uncorrelated wave function, with no Jastrow factor, gives a finite magnetization in a portion of the phase diagram~\cite{supp_mat}).

The inclusion of a sizeable nearest-neighbor Hubbard interaction $V$ induces the onset of charge order. Within our variational approach, we can 
define CDW phases and charge-bond ordered (CBO) phases by suitable choices for the auxiliary Hamiltonian $\mathcal{H}_0$. For CDW states, 
$\mathcal{H}_0$ contains a uniform nearest-neighbor hopping ($T_{i,j}^\sigma=1$) and non-zero onsite potentials $\mu_i$. Instead, CBO states are 
obtained when the hoppings $T_{i,j}^\sigma$ of $\mathcal{H}_0$ take different values on different bonds, breaking the symmetries of the kagome 
lattice. In contrast to renormalization group results~\cite{kiesel2013,wang2013}, our variational phase diagram contains only CDW phases driven 
by the onsite accumulation/depletion of electronic charge. No CBO phases are observed. In order to determine the optimal CDW states, we performed 
several numerical calculations in which the onsite potentials $\mu_i$ have been taken to be periodic over different supercells.
 
The first CDW phase induced by the nearest-neighbor repulsion requires a $2\times 2$ supercell (containing $12$ sites), which breaks the 
transational symmetry of the kagome lattice [blue region in Fig.~\ref{fig:phasediag}~(a)]. The charge pattern of this CDW, depicted in 
Fig.~\ref{fig:phasediag}~(b), is characterized by hexagons whose vertices show alternating excess ($\langle n_i\rangle >n_F$) and deficiency 
($\langle n_i\rangle <n_F$) of the local electron density, surrounded by triangles with a tiny electronic depletion. Thus, the $2\times 2$ CDW 
state breaks the $D_6$ point group symmetry of the kagome lattice down to $D_3$. This CDW phase is insulating, as testified by the presence of a 
finite gap in the energy spectrum of the auxiliary Hamiltonian $\mathcal{H}_0$. A further confirmation comes from the calculation of the small-$q$ 
behavior of the density-density structure factor $N(\vec{q}\,)$~\cite{capello2005,capello2006,supp_mat}. 

When the nearest-neighbor interaction is further increased, we detect the transition towards a set of different CDWs, which are characterized by 
a common feature, namely they satisfy a {\it triangle rule}~\cite{obrien2010}: each unit cell of the kagome lattice contains one site where 
$\langle n_i \rangle >n_F$ (electron accumulation) and two sites where $\langle n_i \rangle <n_F$ (electron depletion); futhermore, the repulsive 
interaction $V$ selects the extended charge patterns in which electron-rich sites are surrounded only by electron-poor sites at nearest-neighbors. 
Analogous patterns have been observed in the phase diagram of the interacting spinless fermions at $1/3$ filling~\cite{wen2010}. Here, the CDW
states satisfying the {\it triangle rule} can be visualized as hard-core dimer configurations on the honeycomb lattice, which is formed by the 
centers of corner-sharing triangular plaquettes~\cite{obrien2010}. Three relevant charge patterns of this kind, dubbed CDW I, II and III, are 
shown in Fig.~\ref{fig:phasediag}~(c). For large values of $V$, the CDW III order with a $\sqrt{3}\times\sqrt{3}$ supercell turns out to be the 
ground state of the system [red region of Fig.~\ref{fig:phasediag}~(a)]. For intermediate values of $V/t$, sandwhiched between the $2\times 2$ 
CDW and the CDW III phases, we find a region in the phase diagram where the lowest variational energy is given by the CDW I and II orders [green 
region in Fig.~\ref{fig:phasediag}~(a)]. Actually, within this phase, not only these two CDWs yield the same variational energy, but they turn out 
to be degenerate with other possible patterns in which CDW I and II order coexist in different portions of the lattice. We remark that all the 
CDWs satisfying the {\it triangle rule} are gapless, as inferred from the energy spectrum of the auxiliary Hamiltonian and the behavior of the 
structure factor $N(\vec{q}\,)$~\cite{supp_mat}.

\begin{figure}
\includegraphics[width=\columnwidth]{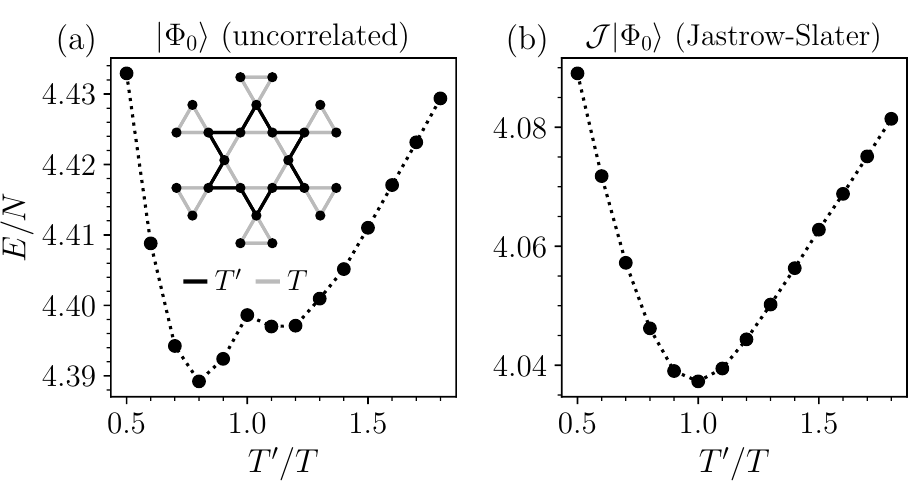}
\caption{\label{fig:cdw2x2landscape}
Variational energy landscape of a wave function reproducing the $2\times 2$ CBO of Ref.~\cite{kiesel2013}, for $U/t=8$ and $V/t=4$. The energy 
is plotted as a function of the ratio between the hopping parameters of $\mathcal{H}_0$ for the nearest-neighbor bonds inside ($T^\prime$) 
and outside ($T$) the star-of-David (shown in the inset). Results for the uncorrelated state $|\Phi_0\rangle$ (a) and the correlated one 
$\mathcal{J}|\Phi_0\rangle$ (b) are shown. The errorbars are smaller than the size of the symbols. The calculations have been performed on a 
finite lattice with $L_1=12$ and $L_2=10$~\cite{supp_mat}.}
\end{figure}

As previously mentioned, while renormalization group calculations detect the presence of CBO and spin-bond order (SBO) in the phase diagram, 
our variational results show that only CDW phases are stabilized by the presence of the nearest-neighbor interaction. To further investigate the 
possibility of bond order, we consider two variational {\it Ans\"atze} that can reproduce the CBO and SBO phases of Ref.~\cite{kiesel2013}. These
are characterized by a $2\times 2$ supercell and a star-of-David pattern for the bond correlations. For the CBO state, we take two distinct 
(spin-isotropic) hopping parameters within $\mathcal{H}_0$, $T^\prime>0$ and $T>0$, for the nearest-neighbor bonds inside and outside the 
star-of-David sketched in Fig.~\ref{fig:cdw2x2landscape}~(a). Interestingly, CBO order is present when the uncorrelated wave function is employed, 
with the energy landscape having a minimum for $T^\prime \ne T$ [Fig.~\ref{fig:cdw2x2landscape} (a)]. However, when the Jastrow factor is included 
to insert electron correlation, the minimum shifts to $T^\prime = T$, indicating no CBO [see Fig.~\ref{fig:cdw2x2landscape}~(b)]. A similar 
analysis can be done to study the possible insurgence of SBO, imposing spin-dependent hoppings and taking $T_\uparrow \le T_\downarrow$ 
($T_\uparrow \ge T_\downarrow$) inside (outside) the star-of-David. Also in this case, a finite order is present only in the uncorrelated state, 
while, in presence of the Jastrow factor, no order is present~\cite{supp_mat}.

Finally, we emphasize that we did not detect the presence of chiral charge order in the phase diagram of the model. To describe 
non-trivial orbital currents and chiral order, we considered variational {\it Ans\"atze} in which the auxiliary Hamiltonian $\mathcal{H}_0$ includes 
complex hopping terms, following the patterns discussed in previous works~\cite{denner2021,park2021,feng2021,mertz2022}. Within the range of couplings 
considered in this work, we find that the complex hopping parameters do not provide any improvement of the variational energy, thus 
implying the absence of interaction-driven chiral charge order in the ground state of the model~\cite{supp_mat}.

\begin{figure}
\includegraphics[width=\columnwidth]{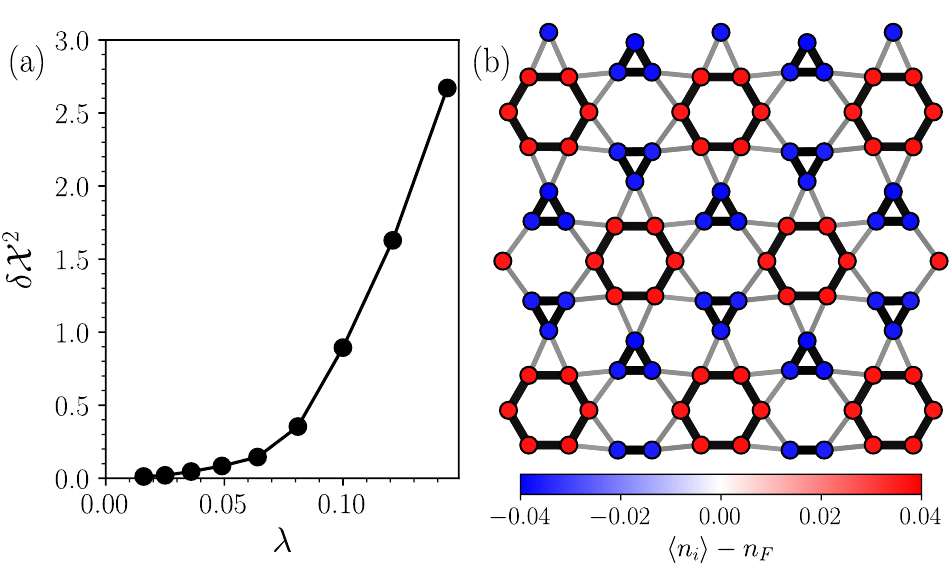}
\caption{\label{fig:chargepattern}
Panel (a): mean squared displacement $\delta \mathcal{X}^2$ as a function of the electron-phonon coupling $\lambda$ in the tri-hexagonal distorted
phase of $\mathcal{H}_{\rm ep}$. Panel (b): lattice distortion induced by the electron-phonon coupling at $\lambda \approx 0.14$. The color of the
sites represents the difference of the local electron density with respect to the filling, $\langle n_i \rangle-n_F$. The width and the darkness 
of $(i,j)$-bonds are proportional to 
$|{\langle c^\dagger_{i,\uparrow} c^{\protect\dagga}_{j,\uparrow} +c^\dagger_{i,\downarrow} c^{\protect\dagga}_{j,\downarrow} \rangle}|$.
The calculations have been performed on a finite lattice with $L_1=8$ and $L_2=6$~\cite{supp_mat}.}
\end{figure}

{\it Electron-phonon coupling.}
The $2\times 2$ CDW that has been obtained within the extended Hubbard model [see Fig.~\ref{fig:phasediag}~(b)] can be hardly reconciled with the 
one observed in AV$_3$Sb$_5$, where the charge show star-of-David and/or tri-hexagonal patterns. This outcome suggests that the origin of the charge 
disproportination in these materials may not be purely electronic and phonons could play an important role. To elucidate this aspect, we consider 
a Su-Schrieffer-Heeger model in which hoppings between electrons are linearly coupled to lattice distortions. The lattice degrees of freedom are 
described by a set of uncoupled harmonic oscillators centered on the lattice sites (Einstein phonons). In addition, the Hubbard-$U$ is considered, 
leading to
\begin{align}\label{eq:ssh_ham}
\mathcal{H}_{\rm ep}&= \sum_{\langle i,j \rangle,\sigma} \left[-t+\alpha \frac{\vec{r}_{i,j}}{||\vec{r}_{i,j}||} \cdot (\vec{u}_i-\vec{u}_j)\right] 
c^\dagger_{i,\sigma} c^\dagga_{j,\sigma} + H.c. \nonumber \\
&+  \sum_i \left(\frac{1}{2m} \vec{p}_i^{\ 2} + \frac{1}{2} m\omega^2 \vec{u}_i^{\ 2} \right)  + U \sum_{i} n_{i,\uparrow} n_{i,\downarrow}.
\end{align}
Here, $\vec{u}_i=(x_i,y_i)$ and $\vec{p}_i=(p^x_i,p^y_i)$ are the displacement and momentum operators of the Einstein phonons. The vectors 
${\vec{r}_{i,j}=(\vec{r}_i-\vec{r}_j)}$ measure the difference of sites positions in the undistorted kagome lattice. The phonon frequency and 
the mass of the ions are denoted by $\omega$ and $m$, respectively, and the strength of the electron-phonon coupling is controlled by the 
parameter $\alpha>0$ or, equivalently, by the dimensionless parameter ${\lambda=\alpha^2/(m\omega^2 t)}$~\cite{stauffert2019}.

We employ a variational wave function that includes both electronic and phononic degrees of freedom~\cite{ferrari2020,ferrari2021}
\begin{equation}
 |\Psi_{\rm ep}\rangle= \mathcal{J}_{\rm ep} |\Psi_{\rm e}\rangle \otimes |\Psi_{\rm p}\rangle,
\end{equation}
where $|\Psi_{\rm e}\rangle$ is the Jastrow-Slater wave function for the electrons, $|\Psi_p\rangle$ is a phonon coherent state, and 
$\mathcal{J}_{\rm ep}$ is an additional Jastrow factor that couples electron and phonon degrees of freedom. For the electron wave function,
we fully optimize the hoppings of the auxiliary Hamiltonian $\mathcal{H}_0$ within a $2\times 2$ supercell. The phonon wave function 
$|\Psi_p\rangle$ is a product of Gaussians in the basis of displacements $\vec{u}_i$
\begin{equation}
\langle \{\vec{u}_i\}|\Psi_p\rangle=\prod_i \exp\left\{-\frac{m\omega}{2 \hbar}\left[(x_i - X_i)^2 + (y_i - Y_i)^2 \right]\right\},
\end{equation}
where $X_i$ and $Y_i$ are variational parameters that control the presence of finite distortions. The electron-phonon Jastrow factor reads
\begin{equation}
 \mathcal{J}_{\rm ep}=\exp\Big \{\sum_{i,j}  n_i n_j  \big[w^x_{i,j}(x_i-x_j) +  w^y_{i,j}  (y_i-y_j)\big] \Big \}.
\end{equation}
The variational parameters $w^x_{i,j}$ and $w^y_{i,j}$ depend only on the distance between lattice sites, $||\vec{r}_{i,j}||$, and are odd under 
the exchange $i \leftrightarrow j$. To compute observables, we use a Monte Carlo approach to sample the infinite Hilbert space of the system in 
the basis of electron occupancies and sites displacements, i.e., $\{|n_i\rangle \otimes |\vec{u}_i\rangle\}$.

The results for $\hbar\omega/t=0.05$ and $U/t=4$, by varying the electron-phonon coupling $\lambda$, are shown in Fig.~\ref{fig:chargepattern}.
We report a measure of the mean squared displacement, $\delta \mathcal{X}^2=1/N \sum_i \langle \Psi_{\rm ep}| {\tilde u}_i^2 |\Psi_{\rm ep}\rangle$ 
(where ${\tilde u}_i^2 = 2m\omega \vec{u}_i^{\,2}/\hbar$), and the local electron density. For large enough $\lambda$, the system develops a tri-hexagonal 
lattice distortion with CBO (also referred to as {\it inverse star-of-David}~\cite{tan2021}), characterized by shrunk hexagons with accumulation 
of electrons, surrounded by shrunk triangles with electron depletion (see Fig.~\ref{fig:chargepattern}). The resulting CDW state is insulating. 
Still, the charge modulation of the distorted phase is in closer agreement with reported scanning tunneling microscopy measurements for 
AV$_{3}$Sb$_{5}$~\cite{jiang2021,zhao2021}, supporting the fact that an electron-phonon mechanism, rather than longer-range electronic repulsions, 
may be at the origin of the the $2\times 2$ charge order observed in these materials. 
 
{\it Conclusions.} We have analyzed the extended Hubbard model on the kagome lattice, with a single orbital on each site, including either the
nearest-neighbor interation $V$ or the electron-phonon coupling $\lambda$, which can be taken as the simplest possible approximation to capture 
some aspects of AV$_{3}$Sb$_{5}$ compounds. Both $V$ and $\lambda$ may stabilize CDW with a $2\times 2$ supercell; however, the charge pattern 
obtained from $V$ is characterized by a sizable $C_6$ to $C_3$ rotational breaking of hexagons, which can be hardly reconciled with experiments on 
AV$_{3}$Sb$_{5}$. A more realistic charge reorganization is found by invoking a phonon mechanism. The present results indicate that CDW formation can 
be described within a minimal model where the multi-orbital character can be neglected. Still, the resulting CDW is insulating 
and no sign of neither interaction-driven nor phonon-driven time-reversal breaking is observed. These facts strongly suggest that the metallic and chiral 
properties have a different origin, which should be ascribed to other degrees of freedom (e.g., Antimony atoms or additional Vanadium orbitals) and 
physical mechanisms (e.g., spin-orbit coupling).

{\it Acknowledgments.}
We are grateful to S. Backes, S. Bhattacharyya, K. Riedl, P. Wunderlich, and R. Thomale for insightful discussions. F.F. acknowledges support 
from the Alexander von Humboldt Foundation through a postdoctoral Humboldt fellowship. F.F. and R.V. acknowledge support by the Deutsche 
Forschungsgemeinschaft (DFG, German Research Foundation) for funding through TRR 288 -- 422213477 (project A05).

\bibliography{biblio}

\begin{thebibliography}{64}
\expandafter\ifx\csname natexlab\endcsname\relax\def\natexlab#1{#1}\fi
\expandafter\ifx\csname bibnamefont\endcsname\relax
  \def\bibnamefont#1{#1}\fi
\expandafter\ifx\csname bibfnamefont\endcsname\relax
  \def\bibfnamefont#1{#1}\fi
\expandafter\ifx\csname citenamefont\endcsname\relax
  \def\citenamefont#1{#1}\fi
\expandafter\ifx\csname url\endcsname\relax
  \def\url#1{\texttt{#1}}\fi
\expandafter\ifx\csname urlprefix\endcsname\relax\def\urlprefix{URL }\fi
\providecommand{\bibinfo}[2]{#2}
\providecommand{\eprint}[2][]{\url{#2}}

\bibitem[{\citenamefont{Hubbard}(1963)}]{hubbard1963}
\bibinfo{author}{\bibfnamefont{J.}~\bibnamefont{Hubbard}},
  \bibinfo{journal}{Proc. Royal Soc. of London} \textbf{\bibinfo{volume}{276}},
  \bibinfo{pages}{238} (\bibinfo{year}{1963}).

\bibitem[{\citenamefont{Bulut et~al.}(2005)\citenamefont{Bulut, Koshibae, and
  Maekawa}}]{bulut2005}
\bibinfo{author}{\bibfnamefont{N.}~\bibnamefont{Bulut}},
  \bibinfo{author}{\bibfnamefont{W.}~\bibnamefont{Koshibae}}, \bibnamefont{and}
  \bibinfo{author}{\bibfnamefont{S.}~\bibnamefont{Maekawa}},
  \bibinfo{journal}{Phys. Rev. Lett.} \textbf{\bibinfo{volume}{95}},
  \bibinfo{pages}{037001} (\bibinfo{year}{2005}),
  \urlprefix\url{https://link.aps.org/doi/10.1103/PhysRevLett.95.037001}.

\bibitem[{\citenamefont{Ohashi et~al.}(2006)\citenamefont{Ohashi, Kawakami, and
  Tsunetsugu}}]{ohashi2006}
\bibinfo{author}{\bibfnamefont{T.}~\bibnamefont{Ohashi}},
  \bibinfo{author}{\bibfnamefont{N.}~\bibnamefont{Kawakami}}, \bibnamefont{and}
  \bibinfo{author}{\bibfnamefont{H.}~\bibnamefont{Tsunetsugu}},
  \bibinfo{journal}{Phys. Rev. Lett.} \textbf{\bibinfo{volume}{97}},
  \bibinfo{pages}{066401} (\bibinfo{year}{2006}),
  \urlprefix\url{https://link.aps.org/doi/10.1103/PhysRevLett.97.066401}.

\bibitem[{\citenamefont{Kuratani et~al.}(2007)\citenamefont{Kuratani, Koga, and
  Kawakami}}]{kuratani2007}
\bibinfo{author}{\bibfnamefont{S.}~\bibnamefont{Kuratani}},
  \bibinfo{author}{\bibfnamefont{A.}~\bibnamefont{Koga}}, \bibnamefont{and}
  \bibinfo{author}{\bibfnamefont{N.}~\bibnamefont{Kawakami}},
  \bibinfo{journal}{Journal of Physics: Condensed Matter}
  \textbf{\bibinfo{volume}{19}}, \bibinfo{pages}{145252}
  (\bibinfo{year}{2007}),
  \urlprefix\url{https://doi.org/10.1088/0953-8984/19/14/145252}.

\bibitem[{\citenamefont{Guertler}(2014)}]{guertler2014}
\bibinfo{author}{\bibfnamefont{S.}~\bibnamefont{Guertler}},
  \bibinfo{journal}{Phys. Rev. B} \textbf{\bibinfo{volume}{90}},
  \bibinfo{pages}{081105} (\bibinfo{year}{2014}),
  \urlprefix\url{https://link.aps.org/doi/10.1103/PhysRevB.90.081105}.

\bibitem[{\citenamefont{Sun and Zhu}(2021)}]{sun2021}
\bibinfo{author}{\bibfnamefont{R.-Y.} \bibnamefont{Sun}} \bibnamefont{and}
  \bibinfo{author}{\bibfnamefont{Z.}~\bibnamefont{Zhu}},
  \bibinfo{journal}{Phys. Rev. B} \textbf{\bibinfo{volume}{104}},
  \bibinfo{pages}{L121118} (\bibinfo{year}{2021}),
  \urlprefix\url{https://link.aps.org/doi/10.1103/PhysRevB.104.L121118}.

\bibitem[{\citenamefont{Kaufmann et~al.}(2021)\citenamefont{Kaufmann, Steiner,
  Scalettar, Held, and Janson}}]{kaufmann2021}
\bibinfo{author}{\bibfnamefont{J.}~\bibnamefont{Kaufmann}},
  \bibinfo{author}{\bibfnamefont{K.}~\bibnamefont{Steiner}},
  \bibinfo{author}{\bibfnamefont{R.}~\bibnamefont{Scalettar}},
  \bibinfo{author}{\bibfnamefont{K.}~\bibnamefont{Held}}, \bibnamefont{and}
  \bibinfo{author}{\bibfnamefont{O.}~\bibnamefont{Janson}},
  \bibinfo{journal}{Phys. Rev. B} \textbf{\bibinfo{volume}{104}},
  \bibinfo{pages}{165127} (\bibinfo{year}{2021}),
  \urlprefix\url{https://link.aps.org/doi/10.1103/PhysRevB.104.165127}.

\bibitem[{\citenamefont{Nishimoto et~al.}(2010)\citenamefont{Nishimoto,
  Nakamura, O'Brien, and Fulde}}]{nishimoto2010}
\bibinfo{author}{\bibfnamefont{S.}~\bibnamefont{Nishimoto}},
  \bibinfo{author}{\bibfnamefont{M.}~\bibnamefont{Nakamura}},
  \bibinfo{author}{\bibfnamefont{A.}~\bibnamefont{O'Brien}}, \bibnamefont{and}
  \bibinfo{author}{\bibfnamefont{P.}~\bibnamefont{Fulde}},
  \bibinfo{journal}{Phys. Rev. Lett.} \textbf{\bibinfo{volume}{104}},
  \bibinfo{pages}{196401} (\bibinfo{year}{2010}),
  \urlprefix\url{https://link.aps.org/doi/10.1103/PhysRevLett.104.196401}.

\bibitem[{\citenamefont{O'Brien et~al.}(2010)\citenamefont{O'Brien, Pollmann,
  and Fulde}}]{obrien2010}
\bibinfo{author}{\bibfnamefont{A.}~\bibnamefont{O'Brien}},
  \bibinfo{author}{\bibfnamefont{F.}~\bibnamefont{Pollmann}}, \bibnamefont{and}
  \bibinfo{author}{\bibfnamefont{P.}~\bibnamefont{Fulde}},
  \bibinfo{journal}{Phys. Rev. B} \textbf{\bibinfo{volume}{81}},
  \bibinfo{pages}{235115} (\bibinfo{year}{2010}),
  \urlprefix\url{https://link.aps.org/doi/10.1103/PhysRevB.81.235115}.

\bibitem[{\citenamefont{Wen et~al.}(2010)\citenamefont{Wen, R\"uegg, Wang, and
  Fiete}}]{wen2010}
\bibinfo{author}{\bibfnamefont{J.}~\bibnamefont{Wen}},
  \bibinfo{author}{\bibfnamefont{A.}~\bibnamefont{R\"uegg}},
  \bibinfo{author}{\bibfnamefont{C.-C.} \bibnamefont{Wang}}, \bibnamefont{and}
  \bibinfo{author}{\bibfnamefont{G.}~\bibnamefont{Fiete}},
  \bibinfo{journal}{Phys. Rev. B} \textbf{\bibinfo{volume}{82}},
  \bibinfo{pages}{075125} (\bibinfo{year}{2010}),
  \urlprefix\url{https://link.aps.org/doi/10.1103/PhysRevB.82.075125}.

\bibitem[{\citenamefont{R\"uegg and Fiete}(2011)}]{ruegg2011}
\bibinfo{author}{\bibfnamefont{A.}~\bibnamefont{R\"uegg}} \bibnamefont{and}
  \bibinfo{author}{\bibfnamefont{G.}~\bibnamefont{Fiete}},
  \bibinfo{journal}{Phys. Rev. B} \textbf{\bibinfo{volume}{83}},
  \bibinfo{pages}{165118} (\bibinfo{year}{2011}),
  \urlprefix\url{https://link.aps.org/doi/10.1103/PhysRevB.83.165118}.

\bibitem[{\citenamefont{Ferhat and Ralko}(2014)}]{ferhat2014}
\bibinfo{author}{\bibfnamefont{K.}~\bibnamefont{Ferhat}} \bibnamefont{and}
  \bibinfo{author}{\bibfnamefont{A.}~\bibnamefont{Ralko}},
  \bibinfo{journal}{Phys. Rev. B} \textbf{\bibinfo{volume}{89}},
  \bibinfo{pages}{155141} (\bibinfo{year}{2014}),
  \urlprefix\url{https://link.aps.org/doi/10.1103/PhysRevB.89.155141}.

\bibitem[{\citenamefont{Yu and Li}(2012)}]{yu2012}
\bibinfo{author}{\bibfnamefont{S.-L.} \bibnamefont{Yu}} \bibnamefont{and}
  \bibinfo{author}{\bibfnamefont{J.-X.} \bibnamefont{Li}},
  \bibinfo{journal}{Phys. Rev. B} \textbf{\bibinfo{volume}{85}},
  \bibinfo{pages}{144402} (\bibinfo{year}{2012}),
  \urlprefix\url{https://link.aps.org/doi/10.1103/PhysRevB.85.144402}.

\bibitem[{\citenamefont{Kiesel and Thomale}(2012)}]{kiesel2012}
\bibinfo{author}{\bibfnamefont{M.}~\bibnamefont{Kiesel}} \bibnamefont{and}
  \bibinfo{author}{\bibfnamefont{R.}~\bibnamefont{Thomale}},
  \bibinfo{journal}{Phys. Rev. B} \textbf{\bibinfo{volume}{86}},
  \bibinfo{pages}{121105} (\bibinfo{year}{2012}),
  \urlprefix\url{https://link.aps.org/doi/10.1103/PhysRevB.86.121105}.

\bibitem[{\citenamefont{Kiesel et~al.}(2013)\citenamefont{Kiesel, Platt, and
  Thomale}}]{kiesel2013}
\bibinfo{author}{\bibfnamefont{M.}~\bibnamefont{Kiesel}},
  \bibinfo{author}{\bibfnamefont{C.}~\bibnamefont{Platt}}, \bibnamefont{and}
  \bibinfo{author}{\bibfnamefont{R.}~\bibnamefont{Thomale}},
  \bibinfo{journal}{Phys. Rev. Lett.} \textbf{\bibinfo{volume}{110}},
  \bibinfo{pages}{126405} (\bibinfo{year}{2013}),
  \urlprefix\url{https://link.aps.org/doi/10.1103/PhysRevLett.110.126405}.

\bibitem[{\citenamefont{Wang et~al.}(2013)\citenamefont{Wang, Li, Xiang, and
  Wang}}]{wang2013}
\bibinfo{author}{\bibfnamefont{W.-S.} \bibnamefont{Wang}},
  \bibinfo{author}{\bibfnamefont{Z.-Z.} \bibnamefont{Li}},
  \bibinfo{author}{\bibfnamefont{Y.-Y.} \bibnamefont{Xiang}}, \bibnamefont{and}
  \bibinfo{author}{\bibfnamefont{Q.-H.} \bibnamefont{Wang}},
  \bibinfo{journal}{Phys. Rev. B} \textbf{\bibinfo{volume}{87}},
  \bibinfo{pages}{115135} (\bibinfo{year}{2013}),
  \urlprefix\url{https://link.aps.org/doi/10.1103/PhysRevB.87.115135}.

\bibitem[{\citenamefont{Ortiz et~al.}(2019)\citenamefont{Ortiz, Gomes, Morey,
  Winiarski, Bordelon, Mangum, Oswald, Rodriguez-Rivera, Neilson, Wilson
  et~al.}}]{ortiz2019}
\bibinfo{author}{\bibfnamefont{B.}~\bibnamefont{Ortiz}},
  \bibinfo{author}{\bibfnamefont{L.}~\bibnamefont{Gomes}},
  \bibinfo{author}{\bibfnamefont{J.}~\bibnamefont{Morey}},
  \bibinfo{author}{\bibfnamefont{M.}~\bibnamefont{Winiarski}},
  \bibinfo{author}{\bibfnamefont{M.}~\bibnamefont{Bordelon}},
  \bibinfo{author}{\bibfnamefont{J.}~\bibnamefont{Mangum}},
  \bibinfo{author}{\bibfnamefont{I.}~\bibnamefont{Oswald}},
  \bibinfo{author}{\bibfnamefont{J.}~\bibnamefont{Rodriguez-Rivera}},
  \bibinfo{author}{\bibfnamefont{J.}~\bibnamefont{Neilson}},
  \bibinfo{author}{\bibfnamefont{S.}~\bibnamefont{Wilson}},
  \bibnamefont{et~al.}, \bibinfo{journal}{Phys. Rev. Materials}
  \textbf{\bibinfo{volume}{3}}, \bibinfo{pages}{094407} (\bibinfo{year}{2019}),
  \urlprefix\url{https://link.aps.org/doi/10.1103/PhysRevMaterials.3.094407}.

\bibitem[{\citenamefont{Jiang et~al.}(2021{\natexlab{a}})\citenamefont{Jiang,
  Wu, Yin, Wang, Hasan, Wilson, Chen, and Hu}}]{jiang2021_rev}
\bibinfo{author}{\bibfnamefont{K.}~\bibnamefont{Jiang}},
  \bibinfo{author}{\bibfnamefont{T.}~\bibnamefont{Wu}},
  \bibinfo{author}{\bibfnamefont{J.-X.} \bibnamefont{Yin}},
  \bibinfo{author}{\bibfnamefont{Z.}~\bibnamefont{Wang}},
  \bibinfo{author}{\bibfnamefont{M.}~\bibnamefont{Hasan}},
  \bibinfo{author}{\bibfnamefont{S.}~\bibnamefont{Wilson}},
  \bibinfo{author}{\bibfnamefont{X.}~\bibnamefont{Chen}}, \bibnamefont{and}
  \bibinfo{author}{\bibfnamefont{J.}~\bibnamefont{Hu}},
  \emph{\bibinfo{title}{Kagome superconductors av$_3$sb$_5$ (a=k, rb, cs)}}
  (\bibinfo{year}{2021}{\natexlab{a}}),
  \urlprefix\url{https://arxiv.org/abs/2109.10809}.

\bibitem[{\citenamefont{Jiang et~al.}(2021{\natexlab{b}})\citenamefont{Jiang,
  Yin, Denner, Shumiya, Ortiz, Xu, Guguchia, He, Hossain, Liu
  et~al.}}]{jiang2021}
\bibinfo{author}{\bibfnamefont{Y.-X.} \bibnamefont{Jiang}},
  \bibinfo{author}{\bibfnamefont{J.-X.} \bibnamefont{Yin}},
  \bibinfo{author}{\bibfnamefont{M.}~\bibnamefont{Denner}},
  \bibinfo{author}{\bibfnamefont{N.}~\bibnamefont{Shumiya}},
  \bibinfo{author}{\bibfnamefont{B.}~\bibnamefont{Ortiz}},
  \bibinfo{author}{\bibfnamefont{G.}~\bibnamefont{Xu}},
  \bibinfo{author}{\bibfnamefont{Z.}~\bibnamefont{Guguchia}},
  \bibinfo{author}{\bibfnamefont{J.}~\bibnamefont{He}},
  \bibinfo{author}{\bibfnamefont{M.}~\bibnamefont{Hossain}},
  \bibinfo{author}{\bibfnamefont{X.}~\bibnamefont{Liu}}, \bibnamefont{et~al.},
  \bibinfo{journal}{Nature Materials} \textbf{\bibinfo{volume}{20}},
  \bibinfo{pages}{1353} (\bibinfo{year}{2021}{\natexlab{b}}),
  \urlprefix\url{https://doi.org/10.1038/s41563-021-01034-y}.

\bibitem[{\citenamefont{Zhao et~al.}(2021)\citenamefont{Zhao, Li, Ortiz,
  Teicher, Park, Ye, Wang, Balents, Wilson, and Zeljkovic}}]{zhao2021}
\bibinfo{author}{\bibfnamefont{H.}~\bibnamefont{Zhao}},
  \bibinfo{author}{\bibfnamefont{H.}~\bibnamefont{Li}},
  \bibinfo{author}{\bibfnamefont{B.}~\bibnamefont{Ortiz}},
  \bibinfo{author}{\bibfnamefont{S.}~\bibnamefont{Teicher}},
  \bibinfo{author}{\bibfnamefont{T.}~\bibnamefont{Park}},
  \bibinfo{author}{\bibfnamefont{M.}~\bibnamefont{Ye}},
  \bibinfo{author}{\bibfnamefont{Z.}~\bibnamefont{Wang}},
  \bibinfo{author}{\bibfnamefont{L.}~\bibnamefont{Balents}},
  \bibinfo{author}{\bibfnamefont{S.}~\bibnamefont{Wilson}}, \bibnamefont{and}
  \bibinfo{author}{\bibfnamefont{I.}~\bibnamefont{Zeljkovic}},
  \bibinfo{journal}{Nature} \textbf{\bibinfo{volume}{599}},
  \bibinfo{pages}{216} (\bibinfo{year}{2021}),
  \urlprefix\url{https://doi.org/10.1038/s41586-021-03946-w}.

\bibitem[{\citenamefont{Uykur et~al.}(2021)\citenamefont{Uykur, Ortiz,
  Iakutkina, Wenzel, Wilson, Dressel, and Tsirlin}}]{uykur2021}
\bibinfo{author}{\bibfnamefont{E.}~\bibnamefont{Uykur}},
  \bibinfo{author}{\bibfnamefont{B.}~\bibnamefont{Ortiz}},
  \bibinfo{author}{\bibfnamefont{O.}~\bibnamefont{Iakutkina}},
  \bibinfo{author}{\bibfnamefont{M.}~\bibnamefont{Wenzel}},
  \bibinfo{author}{\bibfnamefont{S.}~\bibnamefont{Wilson}},
  \bibinfo{author}{\bibfnamefont{M.}~\bibnamefont{Dressel}}, \bibnamefont{and}
  \bibinfo{author}{\bibfnamefont{A.}~\bibnamefont{Tsirlin}},
  \bibinfo{journal}{Phys. Rev. B} \textbf{\bibinfo{volume}{104}},
  \bibinfo{pages}{045130} (\bibinfo{year}{2021}),
  \urlprefix\url{https://link.aps.org/doi/10.1103/PhysRevB.104.045130}.

\bibitem[{\citenamefont{Ortiz et~al.}(2021{\natexlab{a}})\citenamefont{Ortiz,
  Teicher, Kautzsch, Sarte, Ratcliff, Harter, Ruff, Seshadri, and
  Wilson}}]{ortiz2021}
\bibinfo{author}{\bibfnamefont{B.}~\bibnamefont{Ortiz}},
  \bibinfo{author}{\bibfnamefont{S.}~\bibnamefont{Teicher}},
  \bibinfo{author}{\bibfnamefont{L.}~\bibnamefont{Kautzsch}},
  \bibinfo{author}{\bibfnamefont{P.}~\bibnamefont{Sarte}},
  \bibinfo{author}{\bibfnamefont{N.}~\bibnamefont{Ratcliff}},
  \bibinfo{author}{\bibfnamefont{J.}~\bibnamefont{Harter}},
  \bibinfo{author}{\bibfnamefont{J.}~\bibnamefont{Ruff}},
  \bibinfo{author}{\bibfnamefont{R.}~\bibnamefont{Seshadri}}, \bibnamefont{and}
  \bibinfo{author}{\bibfnamefont{S.}~\bibnamefont{Wilson}},
  \bibinfo{journal}{Phys. Rev. X} \textbf{\bibinfo{volume}{11}},
  \bibinfo{pages}{041030} (\bibinfo{year}{2021}{\natexlab{a}}),
  \urlprefix\url{https://link.aps.org/doi/10.1103/PhysRevX.11.041030}.

\bibitem[{\citenamefont{Liang et~al.}(2021)\citenamefont{Liang, Hou, Zhang, Ma,
  Wu, Zhang, Yu, Ying, Jiang, Shan et~al.}}]{liang2021}
\bibinfo{author}{\bibfnamefont{Z.}~\bibnamefont{Liang}},
  \bibinfo{author}{\bibfnamefont{X.}~\bibnamefont{Hou}},
  \bibinfo{author}{\bibfnamefont{F.}~\bibnamefont{Zhang}},
  \bibinfo{author}{\bibfnamefont{W.}~\bibnamefont{Ma}},
  \bibinfo{author}{\bibfnamefont{P.}~\bibnamefont{Wu}},
  \bibinfo{author}{\bibfnamefont{Z.}~\bibnamefont{Zhang}},
  \bibinfo{author}{\bibfnamefont{F.}~\bibnamefont{Yu}},
  \bibinfo{author}{\bibfnamefont{J.-J.} \bibnamefont{Ying}},
  \bibinfo{author}{\bibfnamefont{K.}~\bibnamefont{Jiang}},
  \bibinfo{author}{\bibfnamefont{L.}~\bibnamefont{Shan}}, \bibnamefont{et~al.},
  \bibinfo{journal}{Phys. Rev. X} \textbf{\bibinfo{volume}{11}},
  \bibinfo{pages}{031026} (\bibinfo{year}{2021}),
  \urlprefix\url{https://link.aps.org/doi/10.1103/PhysRevX.11.031026}.

\bibitem[{\citenamefont{Shumiya et~al.}(2021)\citenamefont{Shumiya, Hossain,
  Yin, Jiang, Ortiz, Liu, Shi, Yin, Lei, Zhang et~al.}}]{shumiya2021}
\bibinfo{author}{\bibfnamefont{N.}~\bibnamefont{Shumiya}},
  \bibinfo{author}{\bibfnamefont{M.}~\bibnamefont{Hossain}},
  \bibinfo{author}{\bibfnamefont{J.-X.} \bibnamefont{Yin}},
  \bibinfo{author}{\bibfnamefont{Y.-X.} \bibnamefont{Jiang}},
  \bibinfo{author}{\bibfnamefont{B.}~\bibnamefont{Ortiz}},
  \bibinfo{author}{\bibfnamefont{H.}~\bibnamefont{Liu}},
  \bibinfo{author}{\bibfnamefont{Y.}~\bibnamefont{Shi}},
  \bibinfo{author}{\bibfnamefont{Q.}~\bibnamefont{Yin}},
  \bibinfo{author}{\bibfnamefont{H.}~\bibnamefont{Lei}},
  \bibinfo{author}{\bibfnamefont{S.}~\bibnamefont{Zhang}},
  \bibnamefont{et~al.}, \bibinfo{journal}{Phys. Rev. B}
  \textbf{\bibinfo{volume}{104}}, \bibinfo{pages}{035131}
  (\bibinfo{year}{2021}),
  \urlprefix\url{https://link.aps.org/doi/10.1103/PhysRevB.104.035131}.

\bibitem[{\citenamefont{Li et~al.}(2022)\citenamefont{Li, Zhao, Ortiz, Park,
  Ye, Balents, Wang, Wilson, and Zeljkovic}}]{li2022}
\bibinfo{author}{\bibfnamefont{H.}~\bibnamefont{Li}},
  \bibinfo{author}{\bibfnamefont{H.}~\bibnamefont{Zhao}},
  \bibinfo{author}{\bibfnamefont{B.}~\bibnamefont{Ortiz}},
  \bibinfo{author}{\bibfnamefont{T.}~\bibnamefont{Park}},
  \bibinfo{author}{\bibfnamefont{M.}~\bibnamefont{Ye}},
  \bibinfo{author}{\bibfnamefont{L.}~\bibnamefont{Balents}},
  \bibinfo{author}{\bibfnamefont{Z.}~\bibnamefont{Wang}},
  \bibinfo{author}{\bibfnamefont{S.}~\bibnamefont{Wilson}}, \bibnamefont{and}
  \bibinfo{author}{\bibfnamefont{I.}~\bibnamefont{Zeljkovic}},
  \bibinfo{journal}{Nature Physics} \textbf{\bibinfo{volume}{18}},
  \bibinfo{pages}{265} (\bibinfo{year}{2022}),
  \urlprefix\url{https://doi.org/10.1038/s41567-021-01479-7}.

\bibitem[{\citenamefont{Ortiz et~al.}(2020)\citenamefont{Ortiz, Teicher, Hu,
  Zuo, Sarte, Schueller, Abeykoon, Krogstad, Rosenkranz, Osborn
  et~al.}}]{ortiz2020}
\bibinfo{author}{\bibfnamefont{B.}~\bibnamefont{Ortiz}},
  \bibinfo{author}{\bibfnamefont{S.~L.} \bibnamefont{Teicher}},
  \bibinfo{author}{\bibfnamefont{Y.}~\bibnamefont{Hu}},
  \bibinfo{author}{\bibfnamefont{J.}~\bibnamefont{Zuo}},
  \bibinfo{author}{\bibfnamefont{P.}~\bibnamefont{Sarte}},
  \bibinfo{author}{\bibfnamefont{E.}~\bibnamefont{Schueller}},
  \bibinfo{author}{\bibfnamefont{A.}~\bibnamefont{Abeykoon}},
  \bibinfo{author}{\bibfnamefont{M.}~\bibnamefont{Krogstad}},
  \bibinfo{author}{\bibfnamefont{S.}~\bibnamefont{Rosenkranz}},
  \bibinfo{author}{\bibfnamefont{R.}~\bibnamefont{Osborn}},
  \bibnamefont{et~al.}, \bibinfo{journal}{Phys. Rev. Lett.}
  \textbf{\bibinfo{volume}{125}}, \bibinfo{pages}{247002}
  (\bibinfo{year}{2020}),
  \urlprefix\url{https://link.aps.org/doi/10.1103/PhysRevLett.125.247002}.

\bibitem[{\citenamefont{Ortiz et~al.}(2021{\natexlab{b}})\citenamefont{Ortiz,
  Sarte, Kenney, Graf, Teicher, Seshadri, and Wilson}}]{ortiz2021b}
\bibinfo{author}{\bibfnamefont{B.}~\bibnamefont{Ortiz}},
  \bibinfo{author}{\bibfnamefont{P.}~\bibnamefont{Sarte}},
  \bibinfo{author}{\bibfnamefont{E.}~\bibnamefont{Kenney}},
  \bibinfo{author}{\bibfnamefont{M.}~\bibnamefont{Graf}},
  \bibinfo{author}{\bibfnamefont{S.}~\bibnamefont{Teicher}},
  \bibinfo{author}{\bibfnamefont{R.}~\bibnamefont{Seshadri}}, \bibnamefont{and}
  \bibinfo{author}{\bibfnamefont{S.}~\bibnamefont{Wilson}},
  \bibinfo{journal}{Phys. Rev. Materials} \textbf{\bibinfo{volume}{5}},
  \bibinfo{pages}{034801} (\bibinfo{year}{2021}{\natexlab{b}}),
  \urlprefix\url{https://link.aps.org/doi/10.1103/PhysRevMaterials.5.034801}.

\bibitem[{\citenamefont{Yin et~al.}(2021)\citenamefont{Yin, Tu, Gong, Fu, Yan,
  and Lei}}]{yin2021}
\bibinfo{author}{\bibfnamefont{Q.}~\bibnamefont{Yin}},
  \bibinfo{author}{\bibfnamefont{Z.}~\bibnamefont{Tu}},
  \bibinfo{author}{\bibfnamefont{C.}~\bibnamefont{Gong}},
  \bibinfo{author}{\bibfnamefont{Y.}~\bibnamefont{Fu}},
  \bibinfo{author}{\bibfnamefont{S.}~\bibnamefont{Yan}}, \bibnamefont{and}
  \bibinfo{author}{\bibfnamefont{H.}~\bibnamefont{Lei}},
  \bibinfo{journal}{Chinese Physics Letters} \textbf{\bibinfo{volume}{38}},
  \bibinfo{pages}{037403} (\bibinfo{year}{2021}),
  \urlprefix\url{http://cpl.iphy.ac.cn/EN/abstract/article_105853.shtml}.

\bibitem[{\citenamefont{Chen et~al.}(2021)\citenamefont{Chen, Yang, Hu, Zhao,
  Yuan, Xing, Qian, Huang, Li, Ye et~al.}}]{chen2021}
\bibinfo{author}{\bibfnamefont{H.}~\bibnamefont{Chen}},
  \bibinfo{author}{\bibfnamefont{H.}~\bibnamefont{Yang}},
  \bibinfo{author}{\bibfnamefont{B.}~\bibnamefont{Hu}},
  \bibinfo{author}{\bibfnamefont{Z.}~\bibnamefont{Zhao}},
  \bibinfo{author}{\bibfnamefont{J.}~\bibnamefont{Yuan}},
  \bibinfo{author}{\bibfnamefont{Y.}~\bibnamefont{Xing}},
  \bibinfo{author}{\bibfnamefont{G.}~\bibnamefont{Qian}},
  \bibinfo{author}{\bibfnamefont{Z.}~\bibnamefont{Huang}},
  \bibinfo{author}{\bibfnamefont{G.}~\bibnamefont{Li}},
  \bibinfo{author}{\bibfnamefont{Y.}~\bibnamefont{Ye}}, \bibnamefont{et~al.},
  \bibinfo{journal}{Nature} \textbf{\bibinfo{volume}{599}},
  \bibinfo{pages}{222} (\bibinfo{year}{2021}),
  \urlprefix\url{https://doi.org/10.1038/s41586-021-03983-5}.

\bibitem[{\citenamefont{Christensen et~al.}(2021)\citenamefont{Christensen,
  Birol, Andersen, and Fernandes}}]{christensen2021}
\bibinfo{author}{\bibfnamefont{M.}~\bibnamefont{Christensen}},
  \bibinfo{author}{\bibfnamefont{T.}~\bibnamefont{Birol}},
  \bibinfo{author}{\bibfnamefont{B.}~\bibnamefont{Andersen}}, \bibnamefont{and}
  \bibinfo{author}{\bibfnamefont{R.}~\bibnamefont{Fernandes}},
  \bibinfo{journal}{Phys. Rev. B} \textbf{\bibinfo{volume}{104}},
  \bibinfo{pages}{214513} (\bibinfo{year}{2021}),
  \urlprefix\url{https://link.aps.org/doi/10.1103/PhysRevB.104.214513}.

\bibitem[{\citenamefont{Hu et~al.}(2022)\citenamefont{Hu, Wu, Ortiz, Han,
  Plumb, Wilson, Schnyder, and Shi}}]{hu2022}
\bibinfo{author}{\bibfnamefont{Y.}~\bibnamefont{Hu}},
  \bibinfo{author}{\bibfnamefont{X.}~\bibnamefont{Wu}},
  \bibinfo{author}{\bibfnamefont{B.}~\bibnamefont{Ortiz}},
  \bibinfo{author}{\bibfnamefont{X.}~\bibnamefont{Han}},
  \bibinfo{author}{\bibfnamefont{N.}~\bibnamefont{Plumb}},
  \bibinfo{author}{\bibfnamefont{S.}~\bibnamefont{Wilson}},
  \bibinfo{author}{\bibfnamefont{A.}~\bibnamefont{Schnyder}}, \bibnamefont{and}
  \bibinfo{author}{\bibfnamefont{M.}~\bibnamefont{Shi}},
  \emph{\bibinfo{title}{Coexistence of tri-hexagonal and star-of-david pattern
  in the charge density wave of the kagome superconductor av$_3$sb$_5$}}
  (\bibinfo{year}{2022}), \urlprefix\url{https://arxiv.org/abs/2201.06477}.

\bibitem[{\citenamefont{Yang et~al.}(2020)\citenamefont{Yang, Wang, Ortiz, Liu,
  Gayles, Derunova, Gonzalez-Hernandez, Smejkal, Chen, Parkin
  et~al.}}]{yang2020}
\bibinfo{author}{\bibfnamefont{S.}~\bibnamefont{Yang}},
  \bibinfo{author}{\bibfnamefont{Y.}~\bibnamefont{Wang}},
  \bibinfo{author}{\bibfnamefont{B.}~\bibnamefont{Ortiz}},
  \bibinfo{author}{\bibfnamefont{D.}~\bibnamefont{Liu}},
  \bibinfo{author}{\bibfnamefont{J.}~\bibnamefont{Gayles}},
  \bibinfo{author}{\bibfnamefont{E.}~\bibnamefont{Derunova}},
  \bibinfo{author}{\bibfnamefont{R.}~\bibnamefont{Gonzalez-Hernandez}},
  \bibinfo{author}{\bibfnamefont{L.}~\bibnamefont{Smejkal}},
  \bibinfo{author}{\bibfnamefont{Y.}~\bibnamefont{Chen}},
  \bibinfo{author}{\bibfnamefont{S.}~\bibnamefont{Parkin}},
  \bibnamefont{et~al.}, \bibinfo{journal}{Science Advances}
  \textbf{\bibinfo{volume}{6}}, \bibinfo{pages}{eabb6003}
  (\bibinfo{year}{2020}),
  \urlprefix\url{https://www.science.org/doi/abs/10.1126/sciadv.abb6003}.

\bibitem[{\citenamefont{Wang et~al.}(2021)\citenamefont{Wang, Jiang, Yin, Li,
  Wang, Huang, Shao, Liu, Zhu, Shumiya et~al.}}]{wang2021}
\bibinfo{author}{\bibfnamefont{Z.}~\bibnamefont{Wang}},
  \bibinfo{author}{\bibfnamefont{Y.-X.} \bibnamefont{Jiang}},
  \bibinfo{author}{\bibfnamefont{J.-X.} \bibnamefont{Yin}},
  \bibinfo{author}{\bibfnamefont{Y.}~\bibnamefont{Li}},
  \bibinfo{author}{\bibfnamefont{G.-Y.} \bibnamefont{Wang}},
  \bibinfo{author}{\bibfnamefont{H.-L.} \bibnamefont{Huang}},
  \bibinfo{author}{\bibfnamefont{S.}~\bibnamefont{Shao}},
  \bibinfo{author}{\bibfnamefont{J.}~\bibnamefont{Liu}},
  \bibinfo{author}{\bibfnamefont{P.}~\bibnamefont{Zhu}},
  \bibinfo{author}{\bibfnamefont{N.}~\bibnamefont{Shumiya}},
  \bibnamefont{et~al.}, \bibinfo{journal}{Phys. Rev. B}
  \textbf{\bibinfo{volume}{104}}, \bibinfo{pages}{075148}
  (\bibinfo{year}{2021}),
  \urlprefix\url{https://link.aps.org/doi/10.1103/PhysRevB.104.075148}.

\bibitem[{\citenamefont{Wu et~al.}(2021)\citenamefont{Wu, Wang, Liu, Li, Xu,
  Yin, Gong, Tu, Lei, Dong et~al.}}]{wu2021}
\bibinfo{author}{\bibfnamefont{Q.}~\bibnamefont{Wu}},
  \bibinfo{author}{\bibfnamefont{Z.}~\bibnamefont{Wang}},
  \bibinfo{author}{\bibfnamefont{Q.}~\bibnamefont{Liu}},
  \bibinfo{author}{\bibfnamefont{R.}~\bibnamefont{Li}},
  \bibinfo{author}{\bibfnamefont{S.}~\bibnamefont{Xu}},
  \bibinfo{author}{\bibfnamefont{Q.}~\bibnamefont{Yin}},
  \bibinfo{author}{\bibfnamefont{C.}~\bibnamefont{Gong}},
  \bibinfo{author}{\bibfnamefont{Z.}~\bibnamefont{Tu}},
  \bibinfo{author}{\bibfnamefont{H.}~\bibnamefont{Lei}},
  \bibinfo{author}{\bibfnamefont{T.}~\bibnamefont{Dong}}, \bibnamefont{et~al.},
  \emph{\bibinfo{title}{Revealing the immediate formation of two-fold rotation
  symmetry in charge-density-wave state of kagome superconductor csv$_3$sb$_5$
  by optical polarization rotation measurement}} (\bibinfo{year}{2021}),
  \urlprefix\url{https://arxiv.org/abs/2110.11306}.

\bibitem[{\citenamefont{Yu et~al.}(2021)\citenamefont{Yu, Wang, Zhang, Sander,
  Ni, Lu, Ma, Wang, Zhao, Chen et~al.}}]{yu2021}
\bibinfo{author}{\bibfnamefont{L.}~\bibnamefont{Yu}},
  \bibinfo{author}{\bibfnamefont{C.}~\bibnamefont{Wang}},
  \bibinfo{author}{\bibfnamefont{Y.}~\bibnamefont{Zhang}},
  \bibinfo{author}{\bibfnamefont{M.}~\bibnamefont{Sander}},
  \bibinfo{author}{\bibfnamefont{S.}~\bibnamefont{Ni}},
  \bibinfo{author}{\bibfnamefont{Z.}~\bibnamefont{Lu}},
  \bibinfo{author}{\bibfnamefont{S.}~\bibnamefont{Ma}},
  \bibinfo{author}{\bibfnamefont{Z.}~\bibnamefont{Wang}},
  \bibinfo{author}{\bibfnamefont{Z.}~\bibnamefont{Zhao}},
  \bibinfo{author}{\bibfnamefont{H.}~\bibnamefont{Chen}}, \bibnamefont{et~al.},
  \emph{\bibinfo{title}{Evidence of a hidden flux phase in the topological
  kagome metal csv$_3$sb$_5$}} (\bibinfo{year}{2021}),
  \urlprefix\url{https://arxiv.org/abs/2107.10714}.

\bibitem[{\citenamefont{Mielke et~al.}(2022)\citenamefont{Mielke, Das, Yin,
  Liu, Gupta, Jiang, Medarde, Wu, Lei, Chang et~al.}}]{mielke2022}
\bibinfo{author}{\bibfnamefont{C.}~\bibnamefont{Mielke}},
  \bibinfo{author}{\bibfnamefont{D.}~\bibnamefont{Das}},
  \bibinfo{author}{\bibfnamefont{J.-X.} \bibnamefont{Yin}},
  \bibinfo{author}{\bibfnamefont{H.}~\bibnamefont{Liu}},
  \bibinfo{author}{\bibfnamefont{R.}~\bibnamefont{Gupta}},
  \bibinfo{author}{\bibfnamefont{Y.-X.} \bibnamefont{Jiang}},
  \bibinfo{author}{\bibfnamefont{M.}~\bibnamefont{Medarde}},
  \bibinfo{author}{\bibfnamefont{X.}~\bibnamefont{Wu}},
  \bibinfo{author}{\bibfnamefont{H.~C.} \bibnamefont{Lei}},
  \bibinfo{author}{\bibfnamefont{J.}~\bibnamefont{Chang}},
  \bibnamefont{et~al.}, \bibinfo{journal}{Nature}
  \textbf{\bibinfo{volume}{602}}, \bibinfo{pages}{245} (\bibinfo{year}{2022}),
  \urlprefix\url{https://doi.org/10.1038/s41586-021-04327-z}.

\bibitem[{\citenamefont{Denner et~al.}(2021)\citenamefont{Denner, Thomale, and
  Neupert}}]{denner2021}
\bibinfo{author}{\bibfnamefont{M.}~\bibnamefont{Denner}},
  \bibinfo{author}{\bibfnamefont{R.}~\bibnamefont{Thomale}}, \bibnamefont{and}
  \bibinfo{author}{\bibfnamefont{T.}~\bibnamefont{Neupert}},
  \bibinfo{journal}{Phys. Rev. Lett.} \textbf{\bibinfo{volume}{127}},
  \bibinfo{pages}{217601} (\bibinfo{year}{2021}),
  \urlprefix\url{https://link.aps.org/doi/10.1103/PhysRevLett.127.217601}.

\bibitem[{\citenamefont{Jeong et~al.}(2022)\citenamefont{Jeong, Yang, Kim, Kim,
  Lee, and Han}}]{jeong2022}
\bibinfo{author}{\bibfnamefont{M.}~\bibnamefont{Jeong}},
  \bibinfo{author}{\bibfnamefont{H.-J.} \bibnamefont{Yang}},
  \bibinfo{author}{\bibfnamefont{H.}~\bibnamefont{Kim}},
  \bibinfo{author}{\bibfnamefont{Y.}~\bibnamefont{Kim}},
  \bibinfo{author}{\bibfnamefont{S.}~\bibnamefont{Lee}}, \bibnamefont{and}
  \bibinfo{author}{\bibfnamefont{M.}~\bibnamefont{Han}},
  \emph{\bibinfo{title}{Crucial role of out-of-plane sb-$p$ orbitals in van
  hove singularity formation and electronic correlation for superconducting
  kagome metal csv$_3$sb$_5$}} (\bibinfo{year}{2022}),
  \urlprefix\url{https://arxiv.org/abs/2203.09329}.

\bibitem[{\citenamefont{Park et~al.}(2021)\citenamefont{Park, Ye, and
  Balents}}]{park2021}
\bibinfo{author}{\bibfnamefont{T.}~\bibnamefont{Park}},
  \bibinfo{author}{\bibfnamefont{M.}~\bibnamefont{Ye}}, \bibnamefont{and}
  \bibinfo{author}{\bibfnamefont{L.}~\bibnamefont{Balents}},
  \bibinfo{journal}{Phys. Rev. B} \textbf{\bibinfo{volume}{104}},
  \bibinfo{pages}{035142} (\bibinfo{year}{2021}),
  \urlprefix\url{https://link.aps.org/doi/10.1103/PhysRevB.104.035142}.

\bibitem[{\citenamefont{Lin and Nandkishore}(2021)}]{lin2021}
\bibinfo{author}{\bibfnamefont{Y.-P.} \bibnamefont{Lin}} \bibnamefont{and}
  \bibinfo{author}{\bibfnamefont{R.}~\bibnamefont{Nandkishore}},
  \bibinfo{journal}{Phys. Rev. B} \textbf{\bibinfo{volume}{104}},
  \bibinfo{pages}{045122} (\bibinfo{year}{2021}),
  \urlprefix\url{https://link.aps.org/doi/10.1103/PhysRevB.104.045122}.

\bibitem[{\citenamefont{Feng et~al.}(2021{\natexlab{a}})\citenamefont{Feng,
  Jiang, Wang, and Hu}}]{feng2021}
\bibinfo{author}{\bibfnamefont{X.}~\bibnamefont{Feng}},
  \bibinfo{author}{\bibfnamefont{K.}~\bibnamefont{Jiang}},
  \bibinfo{author}{\bibfnamefont{Z.}~\bibnamefont{Wang}}, \bibnamefont{and}
  \bibinfo{author}{\bibfnamefont{J.}~\bibnamefont{Hu}},
  \bibinfo{journal}{Science Bulletin} \textbf{\bibinfo{volume}{66}},
  \bibinfo{pages}{1384} (\bibinfo{year}{2021}{\natexlab{a}}),
  \urlprefix\url{https://www.sciencedirect.com/science/article/pii/S2095927321003224}.

\bibitem[{\citenamefont{Feng et~al.}(2021{\natexlab{b}})\citenamefont{Feng,
  Zhang, Jiang, and Hu}}]{feng2021b}
\bibinfo{author}{\bibfnamefont{X.}~\bibnamefont{Feng}},
  \bibinfo{author}{\bibfnamefont{Y.}~\bibnamefont{Zhang}},
  \bibinfo{author}{\bibfnamefont{K.}~\bibnamefont{Jiang}}, \bibnamefont{and}
  \bibinfo{author}{\bibfnamefont{J.}~\bibnamefont{Hu}}, \bibinfo{journal}{Phys.
  Rev. B} \textbf{\bibinfo{volume}{104}}, \bibinfo{pages}{165136}
  (\bibinfo{year}{2021}{\natexlab{b}}),
  \urlprefix\url{https://link.aps.org/doi/10.1103/PhysRevB.104.165136}.

\bibitem[{\citenamefont{Mertz et~al.}(2022)\citenamefont{Mertz, Wunderlich,
  Bhattacharyya, Ferrari, and Valent{\'i}}}]{mertz2022}
\bibinfo{author}{\bibfnamefont{T.}~\bibnamefont{Mertz}},
  \bibinfo{author}{\bibfnamefont{P.}~\bibnamefont{Wunderlich}},
  \bibinfo{author}{\bibfnamefont{S.}~\bibnamefont{Bhattacharyya}},
  \bibinfo{author}{\bibfnamefont{F.}~\bibnamefont{Ferrari}}, \bibnamefont{and}
  \bibinfo{author}{\bibfnamefont{R.}~\bibnamefont{Valent{\'i}}},
  \bibinfo{journal}{npj Computational Materials} \textbf{\bibinfo{volume}{8}},
  \bibinfo{pages}{66} (\bibinfo{year}{2022}), ISSN \bibinfo{issn}{2057-3960},
  \urlprefix\url{https://doi.org/10.1038/s41524-022-00745-3}.

\bibitem[{\citenamefont{Yang et~al.}(2022)\citenamefont{Yang, Kim, Jeong, Kim,
  Han, and Lee}}]{yang2022}
\bibinfo{author}{\bibfnamefont{H.-J.} \bibnamefont{Yang}},
  \bibinfo{author}{\bibfnamefont{H.}~\bibnamefont{Kim}},
  \bibinfo{author}{\bibfnamefont{M.}~\bibnamefont{Jeong}},
  \bibinfo{author}{\bibfnamefont{Y.}~\bibnamefont{Kim}},
  \bibinfo{author}{\bibfnamefont{M.}~\bibnamefont{Han}}, \bibnamefont{and}
  \bibinfo{author}{\bibfnamefont{S.}~\bibnamefont{Lee}},
  \emph{\bibinfo{title}{Intertwining orbital current order and
  superconductivity in kagome metal}} (\bibinfo{year}{2022}),
  \urlprefix\url{https://arxiv.org/abs/2203.07365}.

\bibitem[{\citenamefont{Wenzel et~al.}(2021)\citenamefont{Wenzel, Ortiz,
  Wilson, Dressel, Tsirlin, and Uykur}}]{wenzel2021}
\bibinfo{author}{\bibfnamefont{M.}~\bibnamefont{Wenzel}},
  \bibinfo{author}{\bibfnamefont{B.}~\bibnamefont{Ortiz}},
  \bibinfo{author}{\bibfnamefont{S.}~\bibnamefont{Wilson}},
  \bibinfo{author}{\bibfnamefont{M.}~\bibnamefont{Dressel}},
  \bibinfo{author}{\bibfnamefont{A.}~\bibnamefont{Tsirlin}}, \bibnamefont{and}
  \bibinfo{author}{\bibfnamefont{E.}~\bibnamefont{Uykur}},
  \emph{\bibinfo{title}{Optical investigations of rbv$_3$sb$_5$: Multiple
  density-wave gaps and phonon anomalies}} (\bibinfo{year}{2021}),
  \urlprefix\url{https://arxiv.org/abs/2112.07501}.

\bibitem[{\citenamefont{Xie et~al.}(2021)\citenamefont{Xie, Li, Bourges,
  Ivanov, Ye, Yin, Hasan, Luo, Yao, Wang et~al.}}]{xie2021}
\bibinfo{author}{\bibfnamefont{Y.}~\bibnamefont{Xie}},
  \bibinfo{author}{\bibfnamefont{Y.}~\bibnamefont{Li}},
  \bibinfo{author}{\bibfnamefont{P.}~\bibnamefont{Bourges}},
  \bibinfo{author}{\bibfnamefont{A.}~\bibnamefont{Ivanov}},
  \bibinfo{author}{\bibfnamefont{Z.}~\bibnamefont{Ye}},
  \bibinfo{author}{\bibfnamefont{J.-X.} \bibnamefont{Yin}},
  \bibinfo{author}{\bibfnamefont{M.}~\bibnamefont{Hasan}},
  \bibinfo{author}{\bibfnamefont{A.}~\bibnamefont{Luo}},
  \bibinfo{author}{\bibfnamefont{Y.}~\bibnamefont{Yao}},
  \bibinfo{author}{\bibfnamefont{Z.}~\bibnamefont{Wang}}, \bibnamefont{et~al.},
  \emph{\bibinfo{title}{Electron-phonon coupling in the charge density wave
  state of csv$_3$sb$_5$}} (\bibinfo{year}{2021}),
  \urlprefix\url{https://arxiv.org/abs/2111.00654}.

\bibitem[{\citenamefont{Ratcliff et~al.}(2021)\citenamefont{Ratcliff, Hallett,
  Ortiz, Wilson, and Harter}}]{ratcliff2021}
\bibinfo{author}{\bibfnamefont{N.}~\bibnamefont{Ratcliff}},
  \bibinfo{author}{\bibfnamefont{L.}~\bibnamefont{Hallett}},
  \bibinfo{author}{\bibfnamefont{B.}~\bibnamefont{Ortiz}},
  \bibinfo{author}{\bibfnamefont{S.}~\bibnamefont{Wilson}}, \bibnamefont{and}
  \bibinfo{author}{\bibfnamefont{J.}~\bibnamefont{Harter}},
  \bibinfo{journal}{Phys. Rev. Materials} \textbf{\bibinfo{volume}{5}},
  \bibinfo{pages}{L111801} (\bibinfo{year}{2021}),
  \urlprefix\url{https://link.aps.org/doi/10.1103/PhysRevMaterials.5.L111801}.

\bibitem[{\citenamefont{Tan et~al.}(2021)\citenamefont{Tan, Liu, Wang, and
  Yan}}]{tan2021}
\bibinfo{author}{\bibfnamefont{H.}~\bibnamefont{Tan}},
  \bibinfo{author}{\bibfnamefont{Y.}~\bibnamefont{Liu}},
  \bibinfo{author}{\bibfnamefont{Z.}~\bibnamefont{Wang}}, \bibnamefont{and}
  \bibinfo{author}{\bibfnamefont{B.}~\bibnamefont{Yan}},
  \bibinfo{journal}{Phys. Rev. Lett.} \textbf{\bibinfo{volume}{127}},
  \bibinfo{pages}{046401} (\bibinfo{year}{2021}),
  \urlprefix\url{https://link.aps.org/doi/10.1103/PhysRevLett.127.046401}.

\bibitem[{\citenamefont{Luo et~al.}(2022)\citenamefont{Luo, Gao, Liu, Gu, Wu,
  Yi, Jia, Wu, Luo, Xu et~al.}}]{luo2022}
\bibinfo{author}{\bibfnamefont{H.}~\bibnamefont{Luo}},
  \bibinfo{author}{\bibfnamefont{Q.}~\bibnamefont{Gao}},
  \bibinfo{author}{\bibfnamefont{H.}~\bibnamefont{Liu}},
  \bibinfo{author}{\bibfnamefont{Y.}~\bibnamefont{Gu}},
  \bibinfo{author}{\bibfnamefont{D.}~\bibnamefont{Wu}},
  \bibinfo{author}{\bibfnamefont{C.}~\bibnamefont{Yi}},
  \bibinfo{author}{\bibfnamefont{J.}~\bibnamefont{Jia}},
  \bibinfo{author}{\bibfnamefont{S.}~\bibnamefont{Wu}},
  \bibinfo{author}{\bibfnamefont{X.}~\bibnamefont{Luo}},
  \bibinfo{author}{\bibfnamefont{Y.}~\bibnamefont{Xu}}, \bibnamefont{et~al.},
  \bibinfo{journal}{Nature Communications} \textbf{\bibinfo{volume}{13}},
  \bibinfo{pages}{273} (\bibinfo{year}{2022}),
  \urlprefix\url{https://doi.org/10.1038/s41467-021-27946-6}.

\bibitem[{\citenamefont{Wu et~al.}(2022)\citenamefont{Wu, Ortiz, Tan, Wilson,
  Yan, Birol, and Blumberg}}]{wu2022}
\bibinfo{author}{\bibfnamefont{S.}~\bibnamefont{Wu}},
  \bibinfo{author}{\bibfnamefont{B.}~\bibnamefont{Ortiz}},
  \bibinfo{author}{\bibfnamefont{H.}~\bibnamefont{Tan}},
  \bibinfo{author}{\bibfnamefont{S.}~\bibnamefont{Wilson}},
  \bibinfo{author}{\bibfnamefont{B.}~\bibnamefont{Yan}},
  \bibinfo{author}{\bibfnamefont{T.}~\bibnamefont{Birol}}, \bibnamefont{and}
  \bibinfo{author}{\bibfnamefont{G.}~\bibnamefont{Blumberg}},
  \emph{\bibinfo{title}{Charge density wave order in kagome metal av$_3$sb$_5$
  (a= cs, rb, k)}} (\bibinfo{year}{2022}),
  \urlprefix\url{https://arxiv.org/abs/2201.05188}.

\bibitem[{\citenamefont{Liu et~al.}(2022)\citenamefont{Liu, Ma, He, Li, Tan,
  Liu, Xu, Tang, Watanabe, Taniguchi et~al.}}]{liu2022}
\bibinfo{author}{\bibfnamefont{G.}~\bibnamefont{Liu}},
  \bibinfo{author}{\bibfnamefont{X.}~\bibnamefont{Ma}},
  \bibinfo{author}{\bibfnamefont{K.}~\bibnamefont{He}},
  \bibinfo{author}{\bibfnamefont{Q.}~\bibnamefont{Li}},
  \bibinfo{author}{\bibfnamefont{H.}~\bibnamefont{Tan}},
  \bibinfo{author}{\bibfnamefont{Y.}~\bibnamefont{Liu}},
  \bibinfo{author}{\bibfnamefont{J.}~\bibnamefont{Xu}},
  \bibinfo{author}{\bibfnamefont{W.}~\bibnamefont{Tang}},
  \bibinfo{author}{\bibfnamefont{K.}~\bibnamefont{Watanabe}},
  \bibinfo{author}{\bibfnamefont{T.}~\bibnamefont{Taniguchi}},
  \bibnamefont{et~al.}, \emph{\bibinfo{title}{Observation of anomalous
  amplitude modes in the kagome metal csv$_3$sb$_5$}} (\bibinfo{year}{2022}),
  \urlprefix\url{https://arxiv.org/abs/2201.05330}.

\bibitem[{\citenamefont{Mei et~al.}()\citenamefont{Mei, Ye, and
  Chen}}]{mei2022}
\bibinfo{author}{\bibfnamefont{J.-W.} \bibnamefont{Mei}},
  \bibinfo{author}{\bibfnamefont{F.}~\bibnamefont{Ye}}, \bibnamefont{and}
  \bibinfo{author}{\bibfnamefont{X.}~\bibnamefont{Chen}},
  \emph{\bibinfo{title}{Lattice dynamics in the charge-density-wave metal at a
  van-hove-singularity filling}},
  \urlprefix\url{https://arxiv.org/abs/2204.05216}.

\bibitem[{\citenamefont{Su et~al.}(1979)\citenamefont{Su, Schrieffer, and
  Heeger}}]{su1979}
\bibinfo{author}{\bibfnamefont{W.}~\bibnamefont{Su}},
  \bibinfo{author}{\bibfnamefont{J.}~\bibnamefont{Schrieffer}},
  \bibnamefont{and} \bibinfo{author}{\bibfnamefont{A.}~\bibnamefont{Heeger}},
  \bibinfo{journal}{Phys. Rev. Lett.} \textbf{\bibinfo{volume}{42}},
  \bibinfo{pages}{1698} (\bibinfo{year}{1979}),
  \urlprefix\url{https://link.aps.org/doi/10.1103/PhysRevLett.42.1698}.

\bibitem[{sup()}]{supp_mat}
\bibinfo{note}{The Supplemental Material is available at ...}

\bibitem[{\citenamefont{Capello et~al.}(2005)\citenamefont{Capello, Becca,
  Fabrizio, Sorella, and Tosatti}}]{capello2005}
\bibinfo{author}{\bibfnamefont{M.}~\bibnamefont{Capello}},
  \bibinfo{author}{\bibfnamefont{F.}~\bibnamefont{Becca}},
  \bibinfo{author}{\bibfnamefont{M.}~\bibnamefont{Fabrizio}},
  \bibinfo{author}{\bibfnamefont{S.}~\bibnamefont{Sorella}}, \bibnamefont{and}
  \bibinfo{author}{\bibfnamefont{E.}~\bibnamefont{Tosatti}},
  \bibinfo{journal}{Phys. Rev. Lett.} \textbf{\bibinfo{volume}{94}},
  \bibinfo{pages}{026406} (\bibinfo{year}{2005}),
  \urlprefix\url{https://link.aps.org/doi/10.1103/PhysRevLett.94.026406}.

\bibitem[{\citenamefont{Capello et~al.}(2006)\citenamefont{Capello, Becca,
  Yunoki, and Sorella}}]{capello2006}
\bibinfo{author}{\bibfnamefont{M.}~\bibnamefont{Capello}},
  \bibinfo{author}{\bibfnamefont{F.}~\bibnamefont{Becca}},
  \bibinfo{author}{\bibfnamefont{S.}~\bibnamefont{Yunoki}}, \bibnamefont{and}
  \bibinfo{author}{\bibfnamefont{S.}~\bibnamefont{Sorella}},
  \bibinfo{journal}{Phys. Rev. B} \textbf{\bibinfo{volume}{73}},
  \bibinfo{pages}{245116} (\bibinfo{year}{2006}),
  \urlprefix\url{https://link.aps.org/doi/10.1103/PhysRevB.73.245116}.

\bibitem[{\citenamefont{Tocchio et~al.}(2013)\citenamefont{Tocchio, Lee,
  Jeschke, Valent\'{\i}, and Gros}}]{tocchio2013}
\bibinfo{author}{\bibfnamefont{L.}~\bibnamefont{Tocchio}},
  \bibinfo{author}{\bibfnamefont{H.}~\bibnamefont{Lee}},
  \bibinfo{author}{\bibfnamefont{H.}~\bibnamefont{Jeschke}},
  \bibinfo{author}{\bibfnamefont{R.}~\bibnamefont{Valent\'{\i}}},
  \bibnamefont{and} \bibinfo{author}{\bibfnamefont{C.}~\bibnamefont{Gros}},
  \bibinfo{journal}{Phys. Rev. B} \textbf{\bibinfo{volume}{87}},
  \bibinfo{pages}{045111} (\bibinfo{year}{2013}),
  \urlprefix\url{https://link.aps.org/doi/10.1103/PhysRevB.87.045111}.

\bibitem[{\citenamefont{Tocchio et~al.}(2014)\citenamefont{Tocchio, Gros,
  Zhang, and Eggert}}]{tocchio2014}
\bibinfo{author}{\bibfnamefont{L.}~\bibnamefont{Tocchio}},
  \bibinfo{author}{\bibfnamefont{C.}~\bibnamefont{Gros}},
  \bibinfo{author}{\bibfnamefont{X.-F.} \bibnamefont{Zhang}}, \bibnamefont{and}
  \bibinfo{author}{\bibfnamefont{S.}~\bibnamefont{Eggert}},
  \bibinfo{journal}{Phys. Rev. Lett.} \textbf{\bibinfo{volume}{113}},
  \bibinfo{pages}{246405} (\bibinfo{year}{2014}),
  \urlprefix\url{https://link.aps.org/doi/10.1103/PhysRevLett.113.246405}.

\bibitem[{\citenamefont{Kaneko et~al.}(2016)\citenamefont{Kaneko, Tocchio,
  Valent\'{\i}, and Gros}}]{kaneko2016}
\bibinfo{author}{\bibfnamefont{R.}~\bibnamefont{Kaneko}},
  \bibinfo{author}{\bibfnamefont{L.}~\bibnamefont{Tocchio}},
  \bibinfo{author}{\bibfnamefont{R.}~\bibnamefont{Valent\'{\i}}},
  \bibnamefont{and} \bibinfo{author}{\bibfnamefont{C.}~\bibnamefont{Gros}},
  \bibinfo{journal}{Phys. Rev. B} \textbf{\bibinfo{volume}{94}},
  \bibinfo{pages}{195111} (\bibinfo{year}{2016}),
  \urlprefix\url{https://link.aps.org/doi/10.1103/PhysRevB.94.195111}.

\bibitem[{\citenamefont{Bijelic et~al.}(2018)\citenamefont{Bijelic, Kaneko,
  Gros, and Valent\'{\i}}}]{bijelic2018}
\bibinfo{author}{\bibfnamefont{M.}~\bibnamefont{Bijelic}},
  \bibinfo{author}{\bibfnamefont{R.}~\bibnamefont{Kaneko}},
  \bibinfo{author}{\bibfnamefont{C.}~\bibnamefont{Gros}}, \bibnamefont{and}
  \bibinfo{author}{\bibfnamefont{R.}~\bibnamefont{Valent\'{\i}}},
  \bibinfo{journal}{Phys. Rev. B} \textbf{\bibinfo{volume}{97}},
  \bibinfo{pages}{125142} (\bibinfo{year}{2018}),
  \urlprefix\url{https://link.aps.org/doi/10.1103/PhysRevB.97.125142}.

\bibitem[{\citenamefont{Sorella}(2005)}]{sorella2005}
\bibinfo{author}{\bibfnamefont{S.}~\bibnamefont{Sorella}},
  \bibinfo{journal}{Phys. Rev. B} \textbf{\bibinfo{volume}{71}},
  \bibinfo{pages}{241103} (\bibinfo{year}{2005}),
  \urlprefix\url{https://link.aps.org/doi/10.1103/PhysRevB.71.241103}.

\bibitem[{\citenamefont{Stauffert et~al.}(2019)\citenamefont{Stauffert, Walter,
  Berciu, and Krems}}]{stauffert2019}
\bibinfo{author}{\bibfnamefont{O.}~\bibnamefont{Stauffert}},
  \bibinfo{author}{\bibfnamefont{M.}~\bibnamefont{Walter}},
  \bibinfo{author}{\bibfnamefont{M.}~\bibnamefont{Berciu}}, \bibnamefont{and}
  \bibinfo{author}{\bibfnamefont{R.}~\bibnamefont{Krems}},
  \bibinfo{journal}{Phys. Rev. B} \textbf{\bibinfo{volume}{100}},
  \bibinfo{pages}{235129} (\bibinfo{year}{2019}),
  \urlprefix\url{https://link.aps.org/doi/10.1103/PhysRevB.100.235129}.

\bibitem[{\citenamefont{Ferrari et~al.}(2020)\citenamefont{Ferrari,
  Valent\'{\i}, and Becca}}]{ferrari2020}
\bibinfo{author}{\bibfnamefont{F.}~\bibnamefont{Ferrari}},
  \bibinfo{author}{\bibfnamefont{R.}~\bibnamefont{Valent\'{\i}}},
  \bibnamefont{and} \bibinfo{author}{\bibfnamefont{F.}~\bibnamefont{Becca}},
  \bibinfo{journal}{Phys. Rev. B} \textbf{\bibinfo{volume}{102}},
  \bibinfo{pages}{125149} (\bibinfo{year}{2020}),
  \urlprefix\url{https://link.aps.org/doi/10.1103/PhysRevB.102.125149}.

\bibitem[{\citenamefont{Ferrari et~al.}(2021)\citenamefont{Ferrari,
  Valent\'{\i}, and Becca}}]{ferrari2021}
\bibinfo{author}{\bibfnamefont{F.}~\bibnamefont{Ferrari}},
  \bibinfo{author}{\bibfnamefont{R.}~\bibnamefont{Valent\'{\i}}},
  \bibnamefont{and} \bibinfo{author}{\bibfnamefont{F.}~\bibnamefont{Becca}},
  \bibinfo{journal}{Phys. Rev. B} \textbf{\bibinfo{volume}{104}},
  \bibinfo{pages}{035126} (\bibinfo{year}{2021}),
  \urlprefix\url{https://link.aps.org/doi/10.1103/PhysRevB.104.035126}.

\end{thebibliography}

\clearpage
\widetext
\begin{center}
\textbf{\large Supplemental Material for ``Charge-density waves in kagome-lattice extended Hubbard models at the van Hove filling''} \bigskip \bigskip
\end{center}
\twocolumngrid

\setcounter{equation}{0}
\setcounter{figure}{0}
\setcounter{table}{0}
\setcounter{page}{1}
\makeatletter
\renewcommand{\theequation}{S\arabic{equation}}
\renewcommand{\thefigure}{S\arabic{figure}}
\renewcommand{\thetable}{S\Roman{table}} 

\section{Additional details on the purely electronic calculations}

\begin{figure}[h!]
\includegraphics[width=\columnwidth]{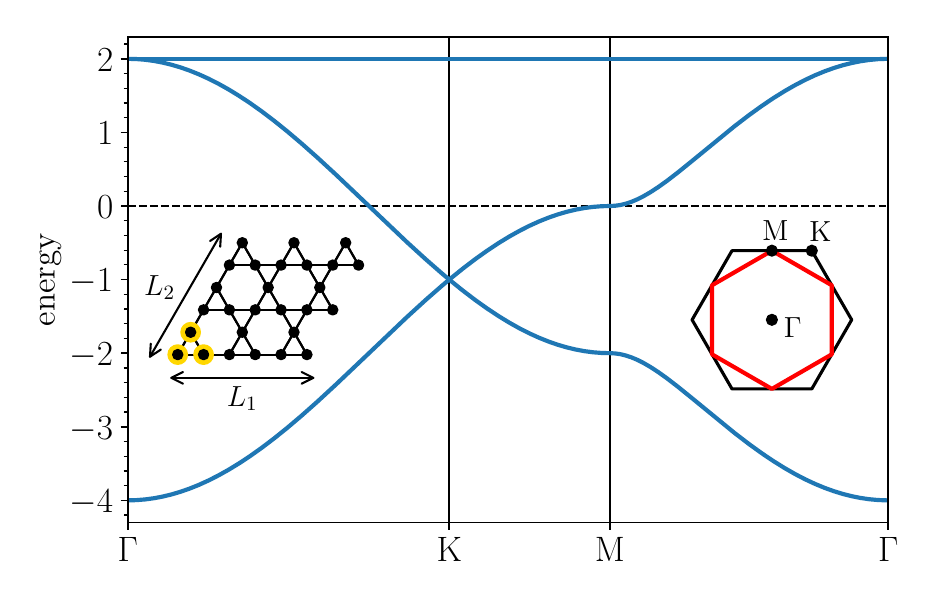}
\caption{
\label{fig:bands}
Band structure of the kagome tight-binding model [Eq.(1) of the main text, with $U=V=0$ and $t=1$]. All bands are doubly degenerate due to spin
degrees of freedom. The horizontal dashed line indicates the Fermi energy at the van Hove filling (i.e., $n_F=5/6$). The left-most inset shows a
$L_1\times L_2\times 3$ kagome lattice, with the three-sites unit cell highlighted in yellow. The right-most inset shows the Brillouin zone of
the lattice (black hexagon) with its high-symmetry points, and the Fermi surface of the tight-binding model at $n_F=5/6$ filling (red hexagon).}
\end{figure}

In Fig.~\ref{fig:bands}, we show the band structure of the single-orbital tight-binding model on the kagome lattice. 
The Fermi energy corresponds to $n_F=5/6$ filling and intersects the upper van Hove singularity.

For our variational calculations we consider finite kagome lattices of $N=L_1 \times L_2 \times 3$ sites, where $L_1$ and $L_2$ indicate the number
of unit cells along the primitive vectors of the underlying triangular Bravais lattice, $\vec{a}_1=(2,0)$ and $\vec{a}_2=(1,\sqrt{3})$.
Most of the results presented in this work are obtained on slightly anisotropic clusters, e.g. with $L_1=12$ and $L_2=10$, and anti-periodic boundary 
conditions for the electrons. This choice ensures the presence of a finite-size gap in the spectrum of the auxiliary Hamiltonian, which allow us to
construct a uniquely defined variational wave function. By performing calculations on different clusters, we have verified that the shape of the 
finite-size lattice does not affect the physical results presented in this work.

\section{Absence of ferromagnetism}

\begin{figure}
\includegraphics[width=\columnwidth]{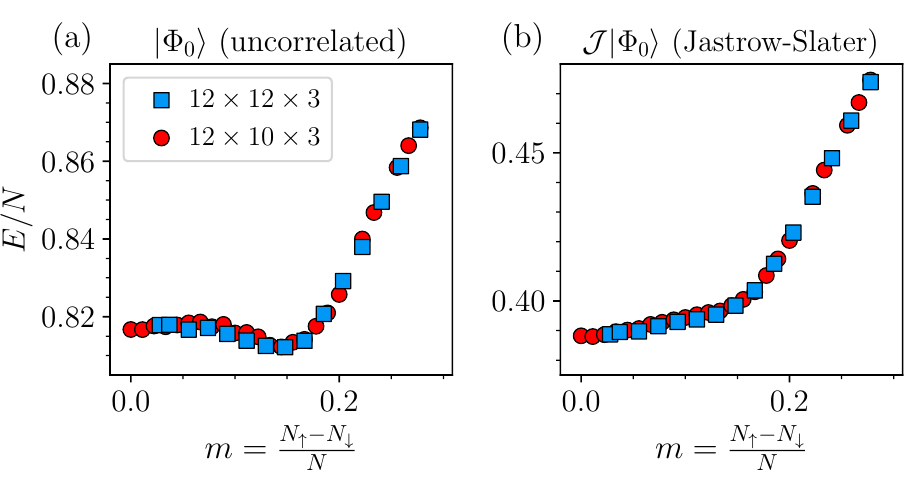}
\caption{\label{fig:ferro}
Variational energy landscape as a function of the magnetization $m$ for $U/t=8$ and $V/t=1$. The energies per site of the uncorrelated wave function
(without Jastrow factor) and the correlated one (with Jastrow factor) are shown in panels (a) and (b), respectively. The error bars are smaller than
the size of the symbols. The calculations have been performed on finite lattices with ${L_1=L_2=12}$ (blue squares) and $L_1=12$ and $L_2=10$ (red
circles).}
\end{figure}

Contrary to renormalization group results~\cite{kiesel2013,wang2013}, for small values of $V/t$ our variational approach indicates the absence of 
ferromagnetism in the ground-state phase diagram of the purely electronic model [Eq.(1) of the main text]. To verify the possible presence of 
ferromagnetism, we compute the variational energy of the metallic state within different sectors of the total $S_z$ operator, i.e., for different 
values of the magnetization $m=(N_\uparrow-N_\downarrow)/N$. For each of these sectors, an independent optimization of the Jastrow parameters is 
performed. In Fig.~\ref{fig:ferro}, we show the variational energy as a function of the magnetization, both for the uncorrelated state 
$|\Phi_0\rangle$ (with no variational parameters) and for the full variational {\it Ansatz} $|\Psi_{\rm e}\rangle=\mathcal{J} |\Phi_0\rangle$. 
Interestingly, the energy landscape of the uncorrelated state shows a minimum at a finite value of the magnetization ($m \approx 0.15$ for $U/t=8$ 
and $V/t=1$), indicating ferromagnetism. However, upon inclusion of correlations through the Jastrow factor, the variational energy largely 
improves, and the minimum of the energy lanscape shifts to zero magnetization. This clearly indicates that the ground state of the system is no
longer ferromagnetic.

\section{Long-wavelength behavior of the density-density structure factor}

\begin{figure}
\includegraphics[width=0.85\columnwidth]{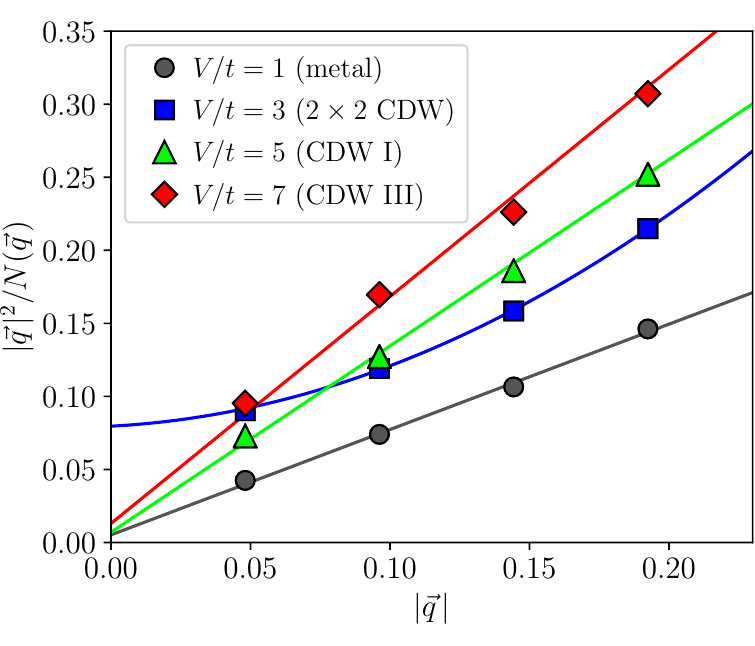}
\caption{\label{fig:nofq}
Inverse of the density-density structure factor $N(\vec{q}\,)$ multiplied by $|\vec{q}\,|^2$, i.e., $|\vec{q}\,|^2/N(\vec{q}\,)$, for small 
values of $|\vec{q}\,|$ along the reciprocal lattice direction of $\vec{b}_1=(\frac{1}{2},-\frac{1}{2\sqrt{3}})$. Results for the electronic 
model [Eq.(1) of the main text] are shown, in the case $U=0$, and for different values of $V/t$. In the limit $|\vec{q}\,|\rightarrow 0$, 
$|\vec{q}\,|^2/N(\vec{q}\,) \rightarrow 0$ linearly vanishes (tends to a finite value) for a metallic (insulating) phase. The lines are fitting 
functions that serve as a guide to the eye.}
\end{figure}

The long-wavelength behavior of the density-density structure factor $N(\vec{q}\,)$ represents a useful tool to detect the presence of a charge 
gap, allowing us to distinguish metallic and insulating phases. The definition of the structure factor is 
\begin{equation}
 N(\vec{q}\,)=\frac{1}{N} \sum_{i,j} e^{i \vec{q} \cdot (\vec{R}_i-\vec{R}_j)} \langle n_i n_j \rangle,
\end{equation}
where $\vec{q}$ is a momentum in the Brillouin zone of the lattice and $\vec{R}_i$ is the coordinate of the unit cell of site $i$. For metallic 
(insulating) phases, $N(\vec{q})\,$ is expected to display a linear (quadratic) behavior in the long-wavelength limit $|\vec{q}\,|\rightarrow 0$,
as discussed in Refs.~\cite{capello2005,capello2006}. In Fig.~\ref{fig:nofq}, we report the quantity $|\vec{q}\,|^2/N(\vec{q}\,)$ at small momenta
$|\vec{q}\,|$, for four values of $V/t$ at $U=0$. In the metallic phases (simple metal, CDW I, II and III) $|\vec{q}\,|^2/N(\vec{q}\,)\rightarrow 0$
for $|\vec{q}\,|\rightarrow 0$; in the insulating $2 \times 2$ CDW phase, instead, $|\vec{q}\,|^2/N(\vec{q}\,)$ tends to a finite value for 
$|\vec{q}\,|\rightarrow 0$, signalling the presence of a non-zero charge gap.

\section{Absence of star-of-David spin-bond order}

\begin{figure}[h!]
\includegraphics[width=\columnwidth]{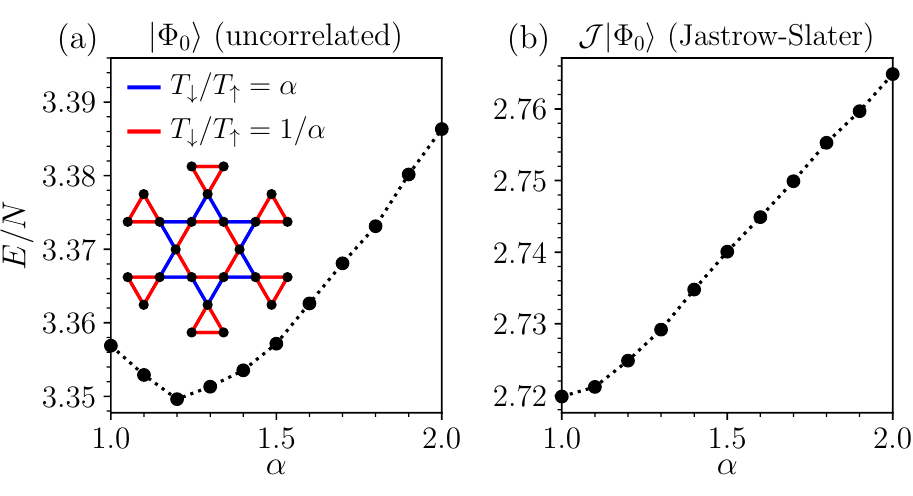}
\caption{\label{fig:sbo2x2landscape}
Variational energy landscape of a wave function reproducing the $2 \times 2$ SBO of Ref.~\cite{kiesel2013}, for the purely electronic model 
[Eq.(1) of the main text] at $U/t=2$ and $V/t=4$. The energy is plotted as a function of the ratio between the spin-dependent hopping parameters 
of the auxiliary Hamiltonian $\mathcal{H}_0$ for the nearest-neighbor bonds inside ($ T_\downarrow/ T_\uparrow=\alpha$) and outside 
($ T_\downarrow/ T_\uparrow=1/\alpha$) the star of David (shown in the inset). Results for the uncorrelated state $|\Phi_0\rangle$ (a) and 
the correlated one $\mathcal{J}|\Phi_0\rangle$ (b) are shown. The errorbars are smaller than the size of the symbols. The calculations have been 
performed on a finite lattice with $L_1=12$ and $L_2=10$.}
\end{figure}

As mentioned in the main text, we verified the possible presence of star-of-David SBO in the phase diagram of the purely electronic system, as 
observed by renormalization group calculations~\cite{kiesel2013}. The SBO wave function is defined by taking a $2 \times 2$ supercell and 
introducing an imbalance between the hoppings for $\uparrow$ and $\downarrow$ spins ($ T_\uparrow, T_\downarrow>0$): for the nearest-neighbor 
bonds inside the star of David sketched in Fig.~\ref{fig:sbo2x2landscape}~(a) we take $ T_\uparrow\le T_\downarrow$, while for those outside 
of it we take $ T_\uparrow \ge  T_\downarrow$ . The results are analogous to those obtained for CBO [Fig.~2 of the main text]: while a 
variational uncorrelated wave function shows presence of SBO, the inclusion of correlations by means of the Jastrow factor leads to a large 
improvement of the variational energy and to the disappearance of SBO (i.e., to $ T_\uparrow= T_\downarrow$). The variational energy landscapes 
for the SBO wave functions are shown in Fig.~\ref{fig:sbo2x2landscape}.

\section{Absence of chiral order}

\begin{figure}[t!]
\includegraphics[width=\columnwidth]{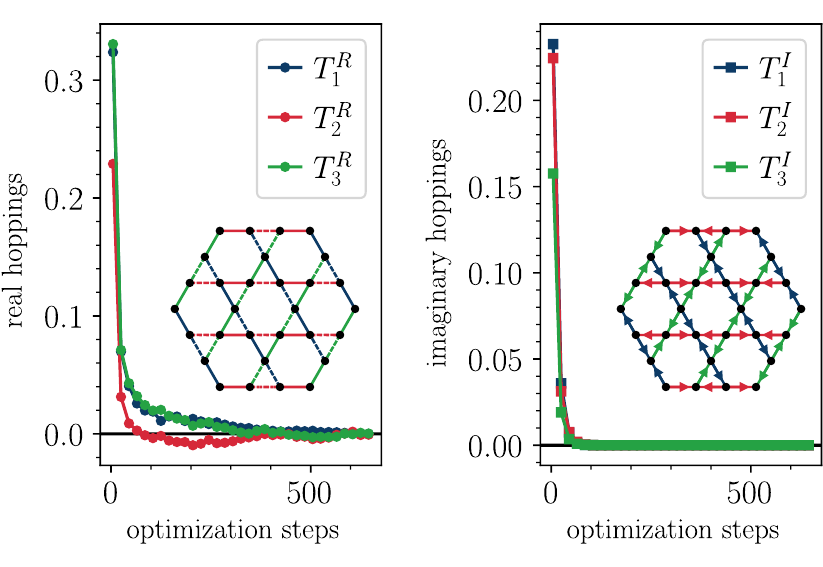}
\caption{\label{fig:chiral}
The evolution of the complex hopping variational parameters upon energy minimization. The results for the real (a) and imaginary (b) parts of 
the hoppings are reported. The calculation is performed for the purely electronic model of Eq.(1) of the main text, at $U/t=4$ and $V/t=3$ (on 
a $L_1=12, \ L_2=10$ cluster). The auxiliary Hamiltonian $\mathcal{H}_0$ of the variational {\it Ansatz} contains a real uniform hopping, which 
is kept fixed to $ T^{\sigma}_{n,m}=1$ during the optimization, and additional complex hoppings (equal for $\uparrow$ and $\downarrow$ spins) 
which reproduce the pattern of the chiral CBO 
order of Ref.~\cite{denner2021}. The patterns of CBO hoppings are shown in the insets, with the different colors (blue, red, and green) denoting 
different hopping parameters ($ T_\alpha^R+i T_\alpha^I$, with $\alpha=1,2$, and $3$). For real hoppings, sketched in the inset of panel (a), 
solid (dashed) lines indicate that the hopping on the bond is $ T_\alpha^R$ ($- T_\alpha^R$). For imaginary hoppings, illustrated in the inset 
of panel (b), an arrow from site $n$ to site $m$ indicates an imaginary hopping term $i  T_\alpha^I \sum_{\sigma} c^\dagger_{n,\sigma}c_{m,\sigma}$
on the bond. Upon optimization, all the chiral CBO hopping terms vanish and the wave function becomes the one of the simple metal.}
\end{figure}

As mentioned in the main text, we did not detect the presence of chiral order and time-reversal breaking in the phase diagram of the electronic 
model. To describe non-trivial orbital currents and chiral order, we have performed variational calculations with an auxiliary Hamiltonian 
$\mathcal{H}_0$ featuring complex hopping terms $ T_{i,j}$. However, the complex parameters do not provide any improvement of the variational 
energy within the region of $U,V$ couplings that we have investigated. As an example, in Fig.~\ref{fig:chiral} we show the flow of the hoppings 
parameters during the optimization of a variational wave function that reproduces the  $2\times 2$ chiral CBO discussed in Ref.~\cite{denner2021}. 
The CBO hoppings of $\mathcal{H}_0$ vanish upon energy minimization and the wave function falls back to the one of the trivial metallic state.

\end{document}